%% file: ao_arxiv.tex
\documentclass[preprint]{emulateapj}

\usepackage{subfigure}
\usepackage{graphicx}
\usepackage[usenames]{color}
\usepackage{amsmath}

\received{}
\revised{}
\accepted{}

\shortauthors{Boyer et al.}
\shorttitle{``Low Mass TP-AGB stars in the Magellanic Clouds''}

\begin{document}

\title{Identification of a Class of Low-Mass Asymptotic Giant Branch Stars Struggling to Become Carbon Stars in the Magellanic Clouds}

\author{Martha~L.~Boyer\altaffilmark{1,2},
  Iain~McDonald\altaffilmark{3},
  Sundar~Srinivasan\altaffilmark{4}
  Albert~Zijlstra\altaffilmark{3},
  Jacco~Th.~van~Loon\altaffilmark{5},
  Knut~A.~G.~Olsen\altaffilmark{6}, and
  George~Sonneborn\altaffilmark{1}}
  \altaffiltext{1}{Observational Cosmology Lab, Code 665, NASA Goddard Space Flight Center, Greenbelt, MD 20771 USA; martha.boyer@nasa.gov}
  \altaffiltext{2}{Oak Ridge Associated Universities (ORAU), Oak Ridge, TN 37831 USA}
  \altaffiltext{3}{Jodrell Bank Centre for Astrophysics, Alan Turing Building, University of Manchester, M13 9PL, UK}
  \altaffiltext{4}{Institute for Astronomy \& Astrophysics, Academia Sinica, 11F, Astronomy-Mathematics Building, No. 1, Roosevelt Rd., Sec 4, Taipei 10617, Taiwan (R.~O.~C.)}
  \altaffiltext{5}{Lennard-Jones Laboratories, Keele University, Staffordshire ST5 5BG, UK}
  \altaffiltext{6}{National Optical Astronomy Observatory, 950 North Cherry Avenue, Tucson, AZ 85719, USA}

\begin{abstract}

\input{abstract_v2.tex}

\end{abstract}


\vfill\eject
\section{INTRODUCTION}
\label{sec:intro}

\subsection{Asymptotic Giant Branch Stars}

Thermally-pulsing Asymptotic Giant Branch (TP-AGB) 
stars contribute substantially to the rest-frame near-infrared (NIR)
luminosities \citep{Maraston+2006,Melbourne+2012,MelbourneBoyer2013}
and to the dust budgets of galaxies
\citep[e.g.,][]{Zhukovska+2013,Schneider+2014,McKinnon+2015}.  The
Magellanic Clouds are excellent galaxies for investigating the global
effects of TP-AGB stars because they are very nearby ($<$65~kpc),
sufficiently massive (0.5--1.5$\times 10^6\,M_\odot^{\rm stars}$), and
have high star-formation rates at epochs required to form TP-AGB
populations in the present day \citep[][and references
  therein]{McConnachie2012}. The TP-AGB stars in both the Large and
Small Magellanic Clouds (LMC/SMC) are therefore extensively studied.
In \citet{Boyer+11} (hereafter B11), we compiled a catalog of TP-AGB
stars in the MCs using photometry from visible, NIR, and IR surveys
and noted a distinct population of TP-AGB stars with
8-\micron\ excesses that separated from the bulk TP-AGB population in
color. We dubbed these the anomalous oxygen-rich (aO)-AGB stars. In
this work, we investigate the nature of these stars using additional
multi-wavelength and time series data.

Before the onset of thermal pulses, stars in the AGB evolutionary
stage are oxygen-rich and their optical to NIR spectra are dominated
by O-rich molecules such as TiO, VO, and H$_2$O. Once thermal pulsing
begins, newly-synthesized elements are dredged-up to the surface. In
intermediate-mass TP-AGB stars, the dredged-up material includes
carbon, which quickly pairs with free oxygen to make CO molecules. If
the dredge-up is efficient enough, the ratio of carbon to oxygen
exceeds unity and the excess carbon is free to form carbon-rich
molecules such as CN and C$_2$. Stellar evolution models
\citep[e.g.,][]{Karakas+02,Marigo+08,Marigo+13} predict a higher
dredge-up efficiency as metallicity decreases. This, combined with
less free oxygen, results in the rapid formation of C stars with fewer
dredge-up events. Carbon stars are therefore more easily formed in
metal-poor environments.

Whether a star remains O-rich or becomes C-rich has drastic
consequences for its subsequent evolution, perhaps affecting the chemical
evolution of its host galaxy.  C-rich TP-AGB stars (C-AGB) easily form
dust and quickly develop strong, radiation-driven winds. O-rich TP-AGB stars (O-AGB) can
also exhibit large mass-loss rates, but in galaxies like the
Magellanic Clouds, it is the C-rich stars that currently dominate the
stellar dust production despite having much smaller numbers than their
O-rich counterparts
\citep{Matsuura+09,Srinivasan+2009,Boyer+2012,Riebel+2012}.
The aO-AGB stars identified in B11 show IR colors intermediate to the
O-AGB and C-AGB population, and their nature was therefore
unclear.

\subsection{Photometric Classifications of TP-AGB stars}

\begin{figure}[h!]
\epsscale{1.1} \vbox{
  \includegraphics[width=\columnwidth]{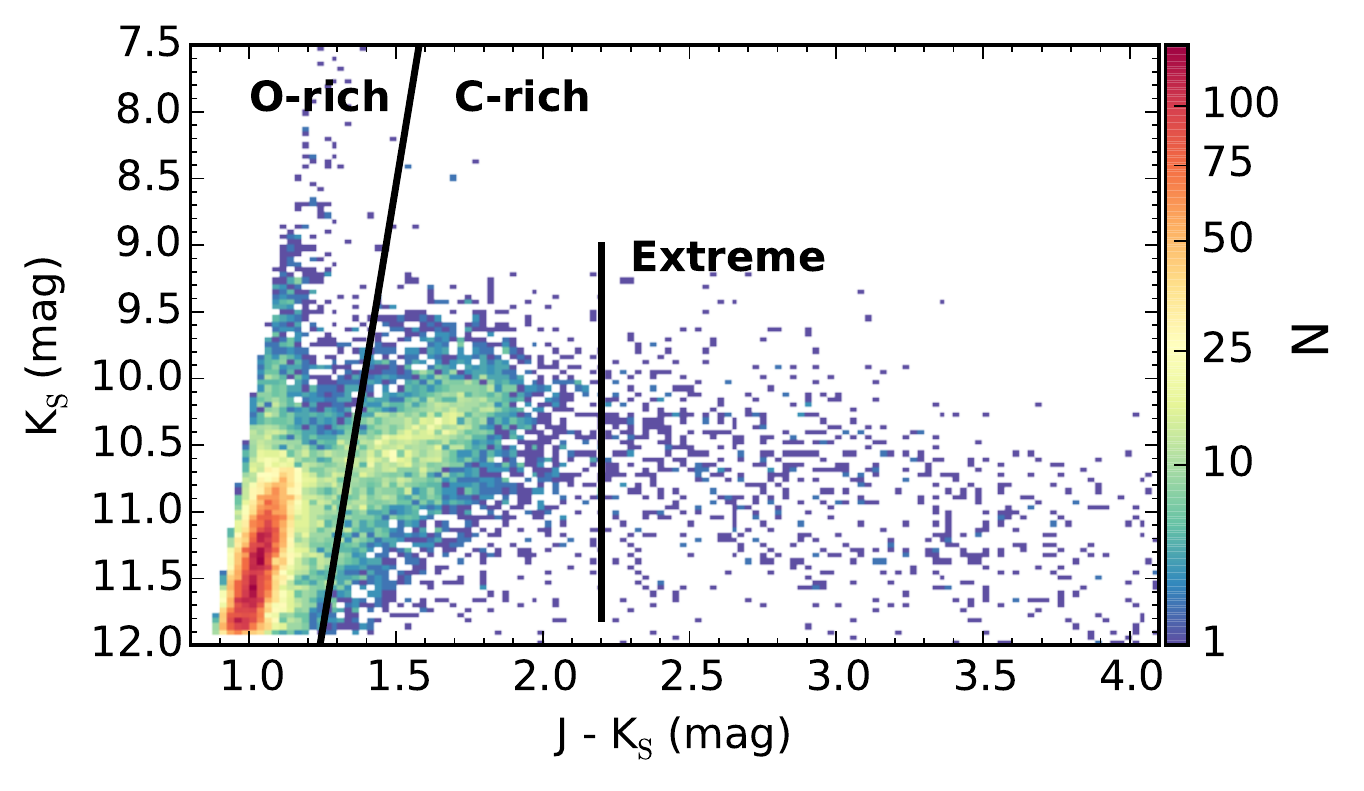}
  \includegraphics[width=\columnwidth]{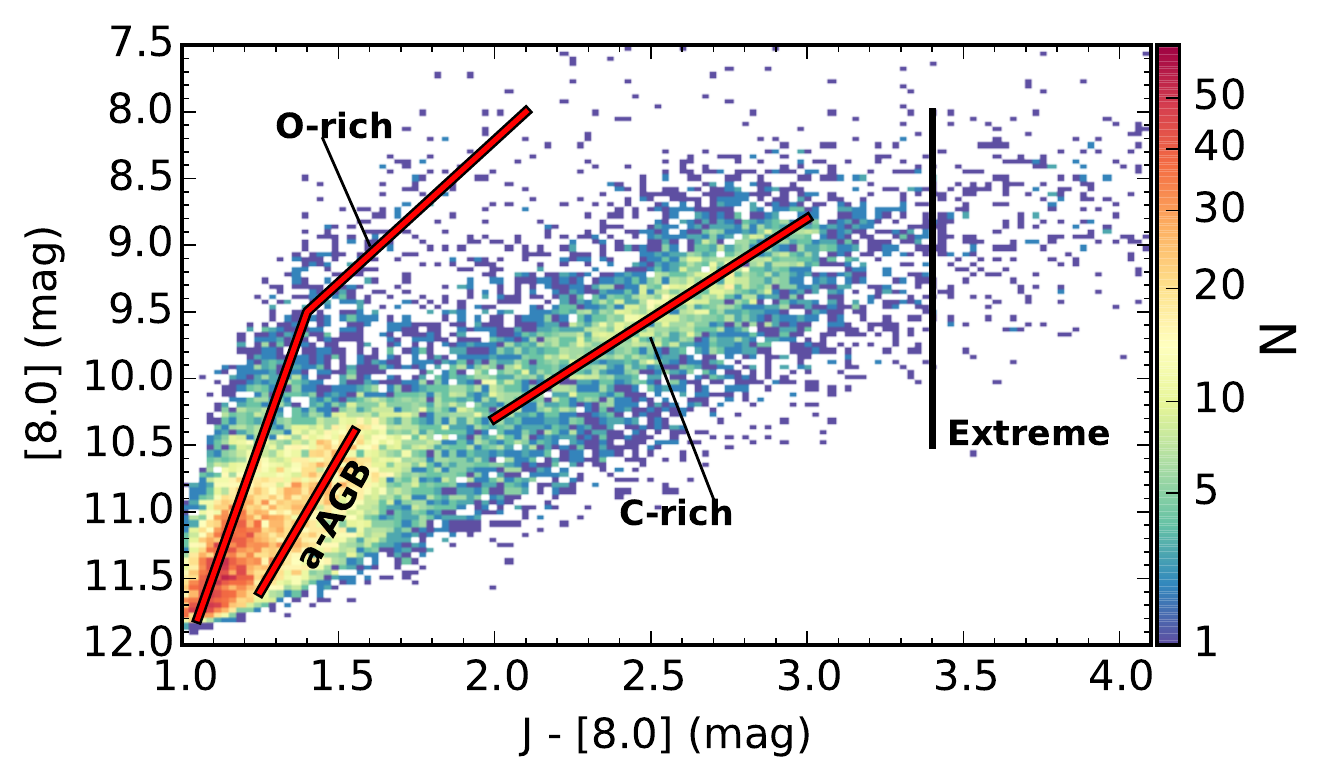} }
\figcaption{Color-magnitude diagrams of LMC TP-AGB stars, as selected
  in B11. {\it Upper Panel:} $J-K_{\rm S}$ versus $K_{\rm s}$ diagram
  showing the \citet{Cioni+06b} and \citet{Blum+06} divisions between
  C-rich, O-rich, and `extreme' (x-)AGB stars (black lines). {\it
    Lower Panel:} $J-[8.0]$ versus $[8.0]$ diagram. The red lines
  highlight the location of O-rich, C-rich, and a-AGB (originally
  named aO-AGB in B11) -- the a-AGB stars form their own sequence in
  this diagram that is separate from the other AGB stars. In both
  panels, the color denotes the density of AGB stars.  In the
  canonical \citet{Cioni+06b} classification scheme, the a-AGB stars
  overlap primarily with O-rich AGB stars, separating only when the
  8-\micron\ photometry is included.\label{fig:photclass}}
\end{figure}

O- and C-rich molecular absorption features are strong at optical and
NIR wavelengths, causing the colors of C-AGB and O-AGB stars to be
drastically different in NIR photometry with commonly used filters.
Several works \citep[e.g.,][]{Cioni+06b} have shown that O-AGB stars
exhibit $J-K_{\rm S} \approx 1$~mag, extending along a mostly vertical
sequence in Figure~\ref{fig:photclass}, upwards from the tip of the
Red Giant Branch (TRGB). C-rich stars follow a more horizontal
sequence in Figure~\ref{fig:photclass}, reaching $J-K_{\rm S} \approx
5$~mag, though very dusty or very cool O-AGB stars \citep[such as
  OH/IR stars and metal-rich stars, e.g.;][]{vanLoon+1998,Boyer+2013}
can also show these red colors.  These molecular features therefore
make NIR photometry an effective, if imperfect, tool for separating
C-AGB and O-AGB stars when spectra are not available.

NIR data from the Two-Micron All-sky Survey
\citep[2MASS;][]{Skrutskie+06} for the Small and Large Magellanic
Clouds (SMC/LMC) show this clear separation between the two types of
AGB stars in $J-K_{\rm S}$ color
\citep[e.g.,][]{Cioni+06b,Blum+06,Boyer+11}. However, B11 found
that, when combined with 8~\micron\ photometry, a third group of stars
distinguishes itself from both the C-AGB and O-AGB stars by showing
$J-[8]$ colors that are redder than the O-AGB stars and bluer than the
C-AGB stars (Fig.~\ref{fig:photclass}). More than 95\% of these stars
are classified as O-rich based on the NIR photometric criteria from
\citet{Cioni+06b}, so we dubbed them anomalous O-rich AGB stars
(aO-AGB).

In B11, we put forth three possible explanations for the seemingly
anomalous $J-[8]$ colors of the aO-AGB stars. First, they may be a
subset of the O-AGB stars that have formed, or are just beginning to
form, dust in their circumstellar envelopes.  Second, these stars may
be a subset of O-AGB stars with particularly strong and/or broad
silicate emission, enhancing their 8-\micron\ flux. A third
possibility is that the aO-AGB stars are S-type AGB stars, which are
transitioning from O-rich to C-rich as they evolve.  It is impossible
to discern whether one of these scenarios is correct with only optical
to mid-IR photometry.  Indeed, we cannot even be sure that the aO-AGB
stars are truly O-rich without additional information.

In this work, we bring together optical spectroscopy, optical to
mid-IR photometry, and optical pulsation information to characterize
the aO-AGB stars with respect to the other O-AGB and C-AGB stars. One
finding, discussed in Section~\ref{sec:specclass}, is that the aO-AGB
stars are not, in fact, all O-rich. We therefore do not wish to
propagate this potentially confusing nomenclature and will instead
refer to the aO-AGB stars simply as a-AGB stars. While the name has
changed, we emphasize that the a-AGB stars are photometrically
classified with a scheme identical to the aO-AGB classification scheme
described by B11 (Section~\ref{sec:photclass}).

With the rich data set compiled here, we compare the chemistry
(Section~\ref{sec:chem}), pulsation (Section~\ref{sec:pulse}), stellar
parameters (Section~\ref{sec:param}), and dust-production
(Section~\ref{sec:mlr}) of the a-AGB stars to the rest of the
Magellanic Cloud TP-AGB stars. All evidence suggests that the a-AGB
stars are low-mass stars at the very end of their evolution. In
Section~\ref{sec:disc}, we compare the results to existing empirical
mass-loss models and to stellar evolution models, and we suggest a
refinement of the canonical TP-AGB classification scheme.

\begin{deluxetable}{lrrr}
\tablewidth{0pc}
\tabletypesize{\small}
\tablecolumns{4}
\tablecaption{Photometry Summary\label{tab:phot}}

\tablehead{\colhead{Catalog} &
\colhead{filter} &
\colhead{Resolution} &
\colhead{Refs.} \\ 
&
\colhead{(\micron\ or band)} & 
\colhead{(\arcsec/pix)} &
} 

\startdata
2MASS & 1.2, 1.6, 2.2& 1\arcsec&  1\\
SAGE & 3.6, 4.5, 5.8, 8.0, 24 & 1\farcs22--2\farcs4& 2,3\\
WISE & 3.4, 4.6, 12, 22& 6\farcs1-12\farcs0 & 4\\
AKARI & 3.2, 4.1, 7, 11, 15, 24 & 2\farcs9--2\farcs8 & 5,6\\
MCPS & 0.37 (U),0.44 (B) & 1\farcs2--1\farcs8 & 7\\
 & 0.55 (V),0.81 (I) & 1\farcs2--1\farcs8 & 7
\enddata
\tablerefs{\ (1) \citet{Skrutskie+06}, (2) \citet{Meixner+06}, (3)
  \citet{Gordon+11}, (4) \citet{Wright+2010}, (5) \citet{Ita+10}, (6)
  \citet{Kato+12}, (7) \citet{Zaritsky+1997}.}
\end{deluxetable}

\section{Data \& Source Classification}
\label{sec:data}

Table~\ref{tab:class} summarizes the final source counts used in our
analysis, classified as described in this section.

\begin{deluxetable}{lrrrr}
\tabletypesize{\normalsize}
\tablecolumns{5}
\tablecaption{Classification Summary\label{tab:class}}

\tablehead{ &
\multicolumn{4}{c}{Photometric Class}\\
&
\colhead{O-AGB} &
\colhead{C-AGB} &
\colhead{x-AGB} &
\colhead{a-AGB}}

\startdata
\multicolumn{5}{l}{SMC: \hrulefill}\\
Total & 2453 & 1708 & 349 & 1266 \\
O-rich spectrum & \nodata & \nodata & \nodata & 100 \\
C-rich spectrum & \nodata & \nodata & \nodata & 122 \\
S-type spectrum & \nodata & \nodata & \nodata & 23 \\
Unknown spectrum & \nodata & \nodata & \nodata & 28 \\
No Spectrum & 2453 & 1708 & 349 & 993 \\
In OGLE-III & 1784 & 1459 & 229 & 1006 \\
\\
\multicolumn{5}{l}{LMC: \hrulefill}\\
Total & 10\,810& 6184 & 1389 & 6342 \\
O-rich spectrum & 1041 & 44 & 4 & 419 \\
C-rich spectrum & 87 & 1501 & 112 & 122 \\
Unknown spectrum & 20 & 23 & 142 & 7 \\
No spectrum & 9662 & 4616 & 1131 & 5794 \\
In OGLE-III & 7445 & 4538 & 710 & 4939 
\enddata
\tablecomments{\ These are the source counts and classifications
  included in our analysis. The columns represent the photometric
  classifications, the rows indicate spectral classifications and
  whether a source is included in the OGLE-III spatial coverage. The
  LMC numbers exclude the kinematically-distinct population
  (Section~\ref{sec:specdata}).}
\end{deluxetable}

\subsection{Photometry}
\label{sec:photdata}

We use the photometry databases compiled by the Surveying the Agents
of Galaxy Evolution (SAGE) Legacy programs
\citep{Meixner+06,Gordon+11}, which includes photometry from the
Magellanic Clouds Photometric Survey \citep[MCPS,
  $UBVI$;][]{Zaritsky+02}, 2MASS \citep[$JHK_{\rm
    s}$;][]{Skrutskie+06}, and {\it Spitzer} (3.6, 4.5, 5.8, 8.0, and
24~\micron) photometry. To this, we add available archival and
literature photometry from the Wide-Field Infrared Survey Explorer
\citep[WISE: 3.4, 4.6, 11.6, and 22~\micron;][]{Wright+2010} and the
      {\it AKARI} targeted surveys of the SMC and LMC \citep[3.2, 4.1,
        7, 11, 15, and
        24~\micron;][]{Ita+10,Kato+12}. Table~\ref{tab:phot}
      summarizes the photometric data used in this study.

The galaxies were covered by {\it Spitzer} twice, 3 months apart. The
photometry used here is derived by combining these two epochs into a
single, co-added stack.  The magnitudes are therefore a flux average
between the two epochs, making them less susceptible to stellar
pulsations.  All magnitudes are corrected for foreground extinction
using the same parameters listed in Table~1 from B11 and the
wavelength dependence derived by \citet{Indebetouw+05}.

\subsubsection{Photometric Classification}
\label{sec:photclass}

A detailed description of the selection criteria for TP-AGB stars is
discussed by B11, and we use identical classifications in this
work. In brief, the NIR $J-K_{\rm S}$ vs. $K_{\rm S}$ color-magnitude
diagram (CMD; Fig.~\ref{fig:photclass}) is used to initially separate
the C-AGB and O-AGB stars, following \citet{Cioni+06b}. We also
classify ``extreme'' dusty AGB stars (x-AGB) as those with $J-[8.0]
\gtrsim 3.4$~mag or $J-K_{\rm S} \gtrsim 2.2$~mag.  Many studies
indicate that the majority of the x-AGB stars in the MCs are C-rich
\citep{Woods+2011,vanLoon+08b,Blum+2014,Ruffle+2015}, though obscured
O-rich stars can also reach very red colors \citep[e.g., the LMC OH/IR
  star IRAS\,05298$-$6957;][]{Wood+1992,vanLoon+01}.  We emphasize
that this classification scheme includes only stars that are brighter
than the 3.6~\micron\ TRGB, which is dominated by AGB stars undergoing
thermal pulsations; we do not include the fainter, less evolved
early-AGB stars since these are difficult to distinguish from red
giant branch stars. Hereafter, we use the term AGB interchangeably
with TP-AGB.

The a-AGB stars are isolated from the initial NIR-selected C- and O-AGB
populations based on their position in the $J-[8]$ vs. $[8]$ CMD
(Fig.~\ref{fig:photclass}). Specifically, the a-AGB sample includes
O-AGB stars redder in $J-[8]$ than the line described by eq.~3 in B11:

\begin{equation}
[8] = A-(11.76 \times (J-[8])),
\end{equation}

\noindent with $A=27.95$, and C-AGB stars bluer than the same line
with $A=31.47$.  The LMC includes $>$6000 a-AGB stars, and they
comprise 26\% of the entire LMC TP-AGB population.  In the SMC, there
are $\sim$1200 a-AGB stars, and they account for a slightly smaller
fraction of the SMC's TP-AGB population (21\%).

These photometric classification are not precise, and there is a small
amount of cross-contamination between each class of stars. We also
stress that the a-AGB and x-AGB classes are defined purely by the
photometric colors; both classes can contain both O-rich and C-rich
AGB stars. In Section~\ref{sec:specclass}, we compare the photometric
classifications to spectral classifications.

\subsection{Variability}
\label{sec:var}

To investigate the pulsation properties of the AGB stars, we include
data from the Optical Gravitational Lensing Experiment
\citep[OGLE-III;][]{Soszynski+08} surveys of the Magellanic
Clouds. The OGLE survey monitored the MCs in the $I$-band for
approximately 8 years, covered the stellar bars in both galaxies, and
derived periods and amplitudes for every variable star.

\begin{figure}[h!]
\epsscale{1.1}
\includegraphics[width=\columnwidth]{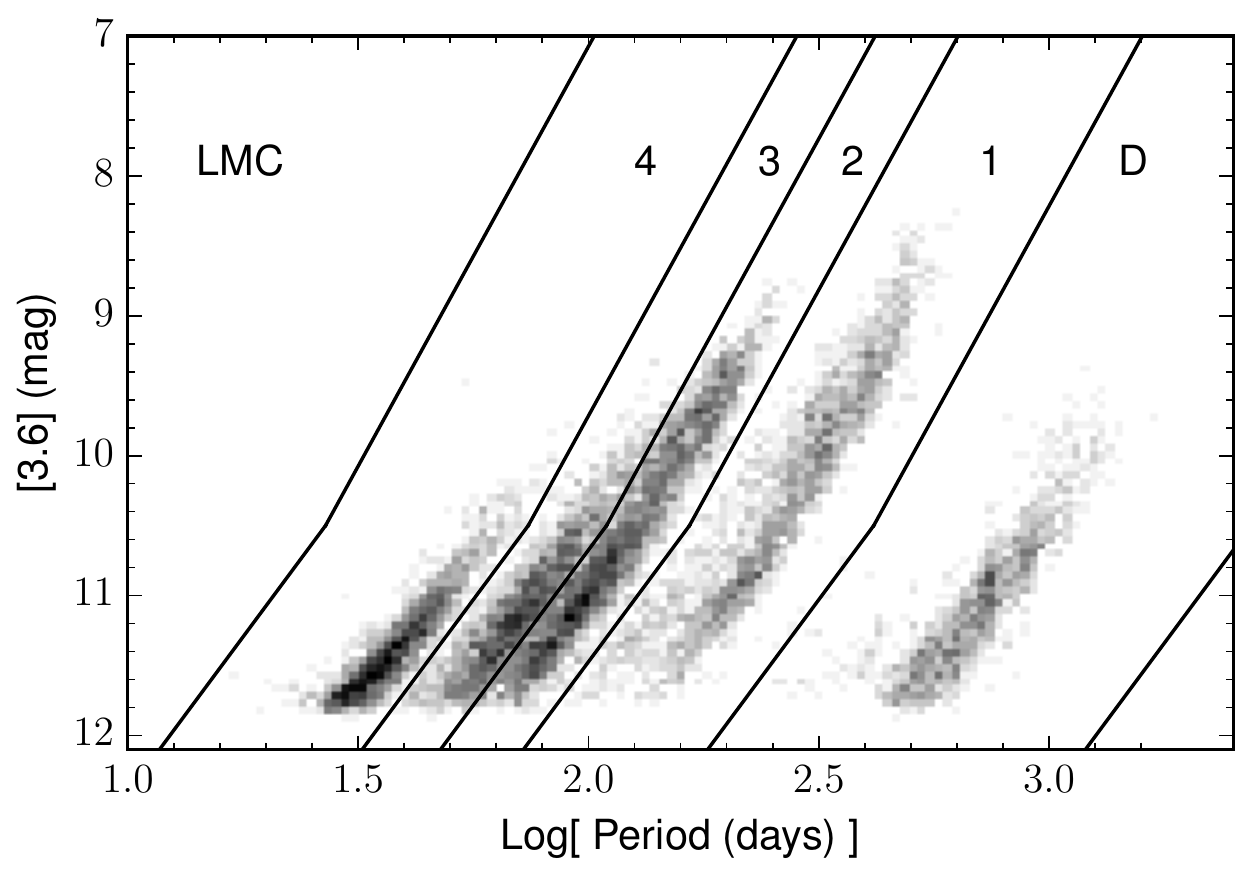}
\figcaption{3.6~\micron\ period-luminosity relationship for the LMC, showing the
  source density. Periods are from the OGLE-III survey and magnitudes are
  from {\it Spitzer}.  Sequences are labeled following
  \citet{Fraser+05}; we assigned stars to a pulsation sequence according to
  the solid lines.  \label{fig:lmc_pl}}
\end{figure}

The 3.6~\micron\ period-luminosity ($P$-$L$) relationship for the LMC
is shown in Figure~\ref{fig:lmc_pl}. The sequences first described by
\citet{Wood+1999} and further subdivided by \citet{Ita+2004} are
clearly evident.  We adopt the \citet{Fraser+05} numbering scheme:
sequences 1--4 describe an evolutionary sequence, with stars evolving
to longer periods and brighter magnitudes as they age. Stars on
Sequence~1 are pulsating in the fundamental mode and higher overtone
pulsators occupy Sequences 2--4. Stars on the upper end of Sequence~1
are often called Mira variables after the prototype Mira,
$o$\,Ceti. The cause of the long secondary periods in stars on
Sequence D is unknown, but might have to do with binarity, nonradial
pulsation, or nonspherical geometry \citep[e.g.,][]{Wood+09}. Here, we
assign stars to each sequence based on the lines drawn in
Figure~\ref{fig:lmc_pl}, with a shift to fainter absolute
3.6-\micron\ magnitudes in the SMC to account for the difference in
distance.

\begin{deluxetable}{llrrr}
\tabletypesize{\normalsize}
\tablecolumns{5}
\tablecaption{Spectroscopic Data Summary\label{tab:data}}

\tablehead{
\colhead{Galaxy} &
\colhead{Instrument} &
\colhead{$\lambda$} &
\colhead{$R$} &
\colhead{Date}}

\startdata
SMC & AAOmega/2dF & 4225--4975\AA & 3700 & Sep 2011\\ 
SMC & AAOmega/2dF & 6325--6775\AA & 8000 & Sep 2011\\
LMC &Hydra-CTIO\tablenotemark{a}  & 6850--9150\AA & 7000 & Nov 2007 

\enddata
\tablenotetext{a}{\ See \citet{Olsen+11} for a detailed description of the Hydra-CTIO observations and data reduction.}
\end{deluxetable}

\subsection{Spectra}
\label{sec:specdata}

We obtained medium-resolution optical spectra of 273 randomly selected
a-AGB stars in the SMC bar using the AAOmega/2dF multi-object
spectrograph \citep{Lewis+02,Saunders+04,Sharp+06} on the
Anglo-Australian Telescope. The total exposure time was 3~hr, and
reductions were carried out with the 2dF pipeline ({\it 2dFdr},
version 4.66, 2011 October). Medium-resolution optical spectra for AGB
stars in the LMC were obtained from 2007 November 20--26 at the CTIO
4~m Blanco telescope with the Hydra-CTIO multi-fiber spectrograph
\citep{Barden+98}. The LMC sample comprises 3791 AGB stars in both the stellar bar and the disk. See \citet{Olsen+11} for
a detailed description of the observations and data
reductions. \citet{Olsen+11} used the LMC spectra to identify a kinematically- and
chemically-distinct population of stars among the LMC sample. They
concluded that this distinct population was accreted from the SMC
during a close passage between the clouds.  We exclude this
  accreted population from the analysis in this paper; 3522 AGB stars,
  including 548 a-AGB stars, remain in our sample
  (Table~\ref{tab:class}).

\begin{figure}[h!]
\epsscale{1.1} \vbox{
  \includegraphics[width=0.9\columnwidth]{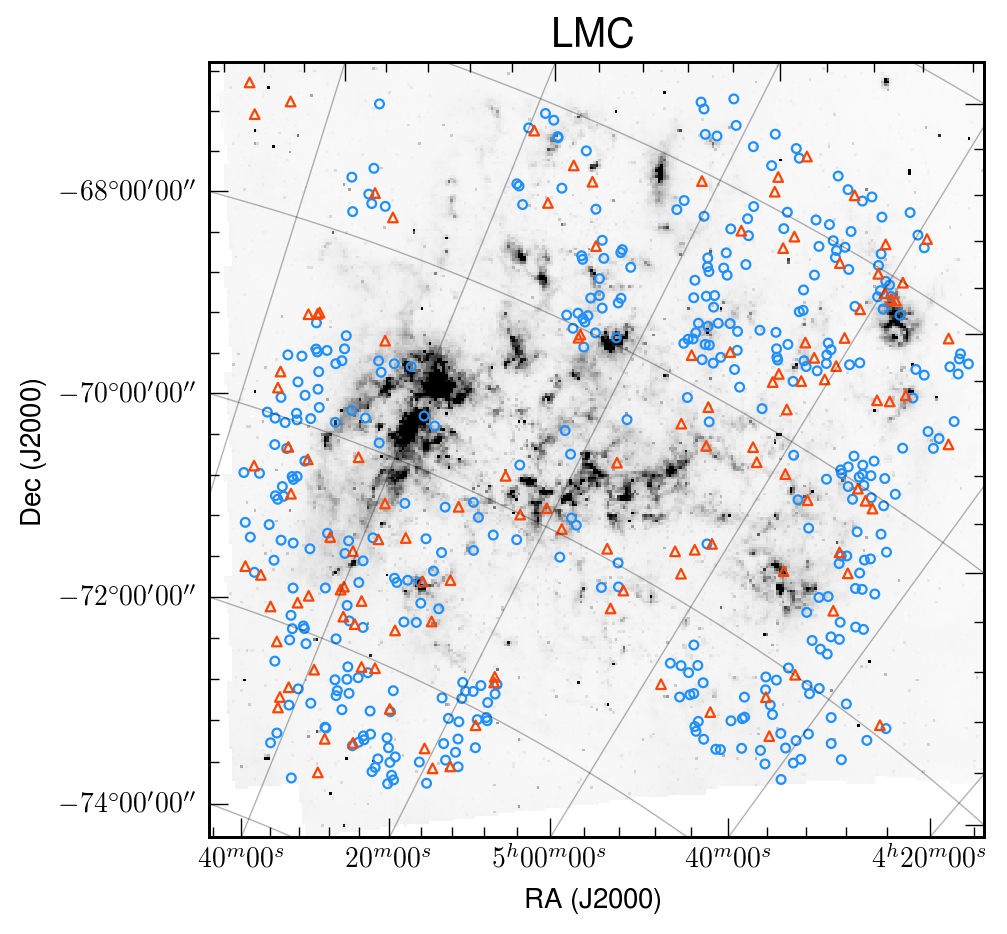}
  \includegraphics[width=0.9\columnwidth]{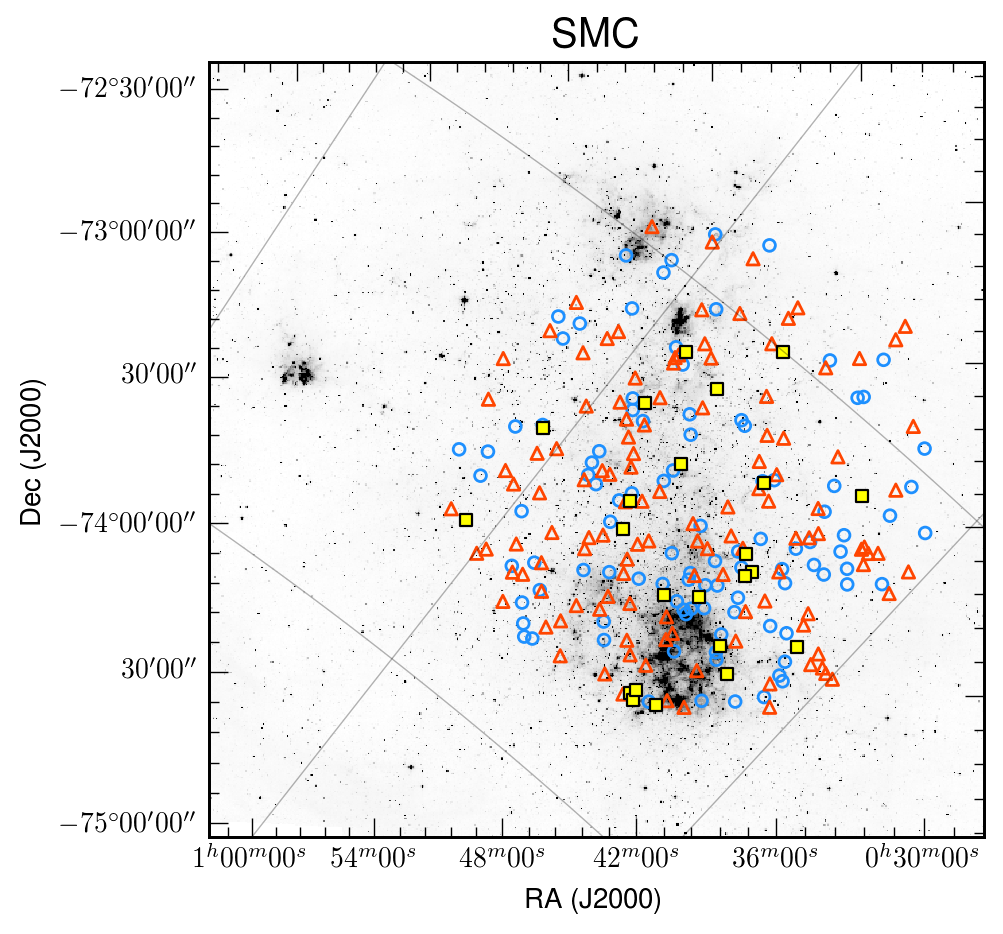}
} \figcaption{Map of the spectrally-classified a-AGB stars in the LMC
  (upper panel) and SMC (lower panel).  The background is the
  8-\micron\ {\it Spitzer} image \citep{Meixner+06,Gordon+11}. The
  a-AGB stars are marked as blue circles (O-rich), red triangles
  (C-rich), and yellow squares (S-type), as classified in
  Section~\ref{sec:specclass}.  \label{fig:map}}
\end{figure}

Spectroscopic data are summarized in Table~\ref{tab:data}.
Figure~\ref{fig:map} shows the map of the a-AGB spectra in both
galaxies, classified as described in the next section.

\begin{figure}[h!]
\epsscale{1.2}
\includegraphics[width=\columnwidth]{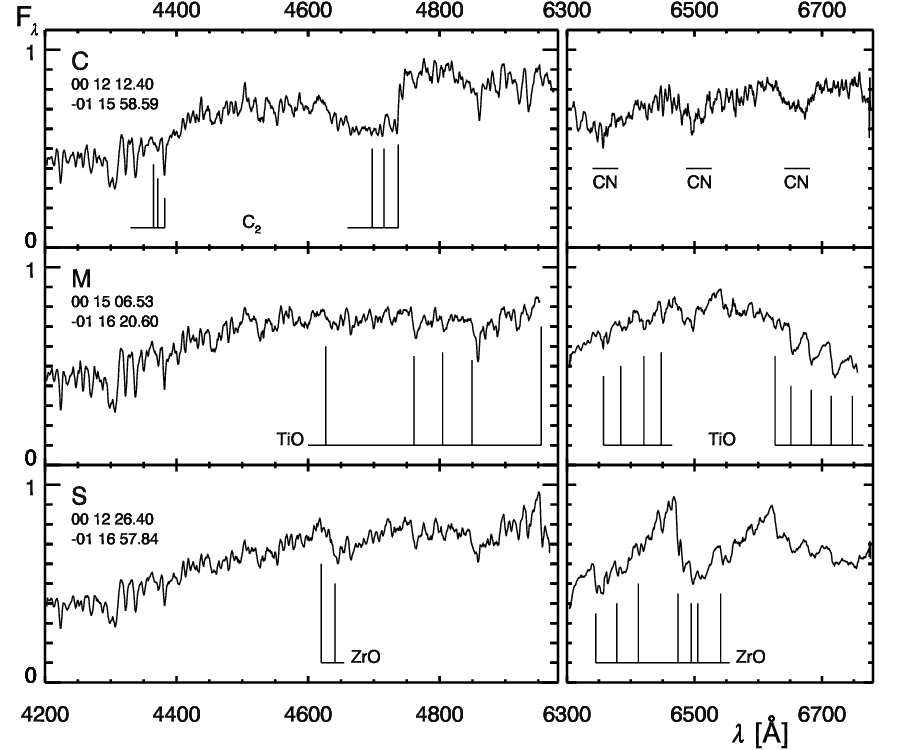}
\figcaption{Example AAOmega/2dF specta of SMC a-AGB stars in
  normalized flux, binned to a resolution of $R \sim 400$. Molecular
  features used to classify stars are marked. C$_2$ and CN bandheads
  are from \citet{Davis87}, TiO from \citet{Valenti+98}, and ZrO from
  \citet{Garcia+07}. \label{fig:aat}}
\end{figure}

\subsubsection{Spectral Classification}
\label{sec:specclass}

The AAOmega/2dF spectra (SMC; a-AGB stars only) are classified by eye
as C (C-rich) or M (O-rich) stars based on the presence of C$_2$$+$CN
or TiO, respectively (Fig.~\ref{fig:aat}). S-type AGB stars are
identified by the presence of ZrO.  This classification scheme results
in 122 C stars, 100 M stars, 23 S stars, and 28 unknown sources
(Table~\ref{tab:class}). We list the SMC S stars in
Table~\ref{tab:Stype}.

\begin{deluxetable}{clrrr}
\tablewidth{0pc}
\tabletypesize{\small}
\tablecolumns{5}
\tablecaption{Spectroscopically-identified S-type Stars in the SMC\label{tab:Stype}}

\tablehead{&\colhead{Designation} &
\colhead{[3.6]} &
\colhead{[8.0]} &
\colhead{Period}\\
&
\colhead{(SSTISAGEMA)} & 
\colhead{(mag)}&
\colhead{(mag)}&
\colhead{(d)}
}

\startdata
 1 & J004234.17-725300.6 & 11.29$\pm$0.01 & 10.92$\pm$0.03 & \nodata\\
 2 & J004428.44-730927.9 & 11.78$\pm$0.02 & 11.70$\pm$0.02 &   80\\
 3 & J004526.04-721432.9 & 11.66$\pm$0.03 & 11.39$\pm$0.03 & \nodata\\
 4 & J004547.21-730535.0 & 11.31$\pm$0.01 & 11.18$\pm$0.01 &  110\\
 5 & J004618.11-732652.4 & 12.44$\pm$0.02 & 12.20$\pm$0.05 &  401\\
 6 & J004712.99-724637.6 & 11.38$\pm$0.01 & 11.03$\pm$0.02 &  653\\
 7 & J004721.34-724834.3 & 11.65$\pm$0.01 & 11.45$\pm$0.02 &   91\\
 8 & J004731.02-732942.6 & 11.20$\pm$0.01 & 11.04$\pm$0.02 &  165\\
 9 & J004745.83-732719.6 & 11.51$\pm$0.01 & 11.33$\pm$0.02 &   95\\
10 & J004755.46-732855.3 & 11.42$\pm$0.01 & 11.36$\pm$0.02 &   82\\
11 & J004808.00-724406.3 & 11.65$\pm$0.01 & 11.45$\pm$0.02 &  685\\
12 & J004833.34-725953.1 & 11.62$\pm$0.02 & 11.38$\pm$0.02 &   88\\
13 & J004955.64-722755.3 & 11.70$\pm$0.03 & 11.49$\pm$0.04 &   89\\
14 & J005005.38-730500.1 & 11.85$\pm$0.01 & 11.65$\pm$0.02 &  116\\
15 & J005342.03-720007.5 & 11.86$\pm$0.03 & 11.73$\pm$0.04 &   99\\
16 & J005404.72-723727.8 & 11.37$\pm$0.01 & 11.27$\pm$0.02 &   89\\
17 & J005418.85-725905.5 & 11.37$\pm$0.02 & 11.25$\pm$0.02 &  105\\
18 & J005458.13-725233.2 & 11.90$\pm$0.02 & 11.67$\pm$0.02 &  659\\
19 & J005510.79-721734.0 & 11.44$\pm$0.01 & 11.33$\pm$0.02 &  109\\
20 & J005744.57-721509.1 & 10.66$\pm$0.01 & 10.52$\pm$0.02 &  123\\
21 & J005746.46-723130.4 & 11.76$\pm$0.01 & 11.58$\pm$0.02 &   93\\
22 & J010119.74-725149.3 & 11.62$\pm$0.02 & 11.44$\pm$0.02 &  606\\
23 & J010133.86-732135.8 & 11.61$\pm$0.03 & 11.41$\pm$0.04 &   92
\enddata

\tablecomments{\ The full photometric catalog is available online,
  including $UBVI$ from MCPS, $JHK_{\rm S}$ from 2MASS, and IRAC and
  MIPS magnitudes from SAGE (see Table~\ref{tab:phot} for references). The OGLE-III
    periods and amplitudes are also included in the online version. }
\end{deluxetable}

\begin{figure}[h!]
\epsscale{1.2}
\includegraphics[width=\columnwidth]{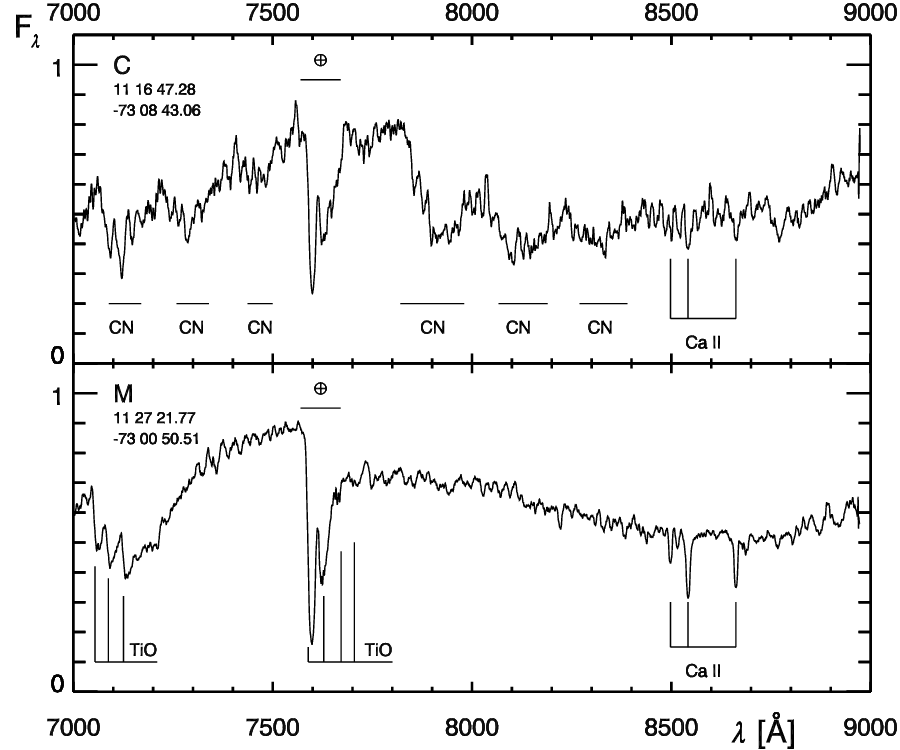}
\figcaption{ Example CTIO specta (in normalized flux) of a C and M AGB
  star in the LMC, binned to $R\sim400$ \citep{Olsen+11}. Molecular
  features used to classify stars are marked. CN bandheads are from
  \citet{Davis87}, TiO from \citet{Valenti+98}. \label{fig:knut}}
\end{figure}

\begin{figure}[h!]
\epsscale{1.1} \includegraphics[width=\columnwidth]{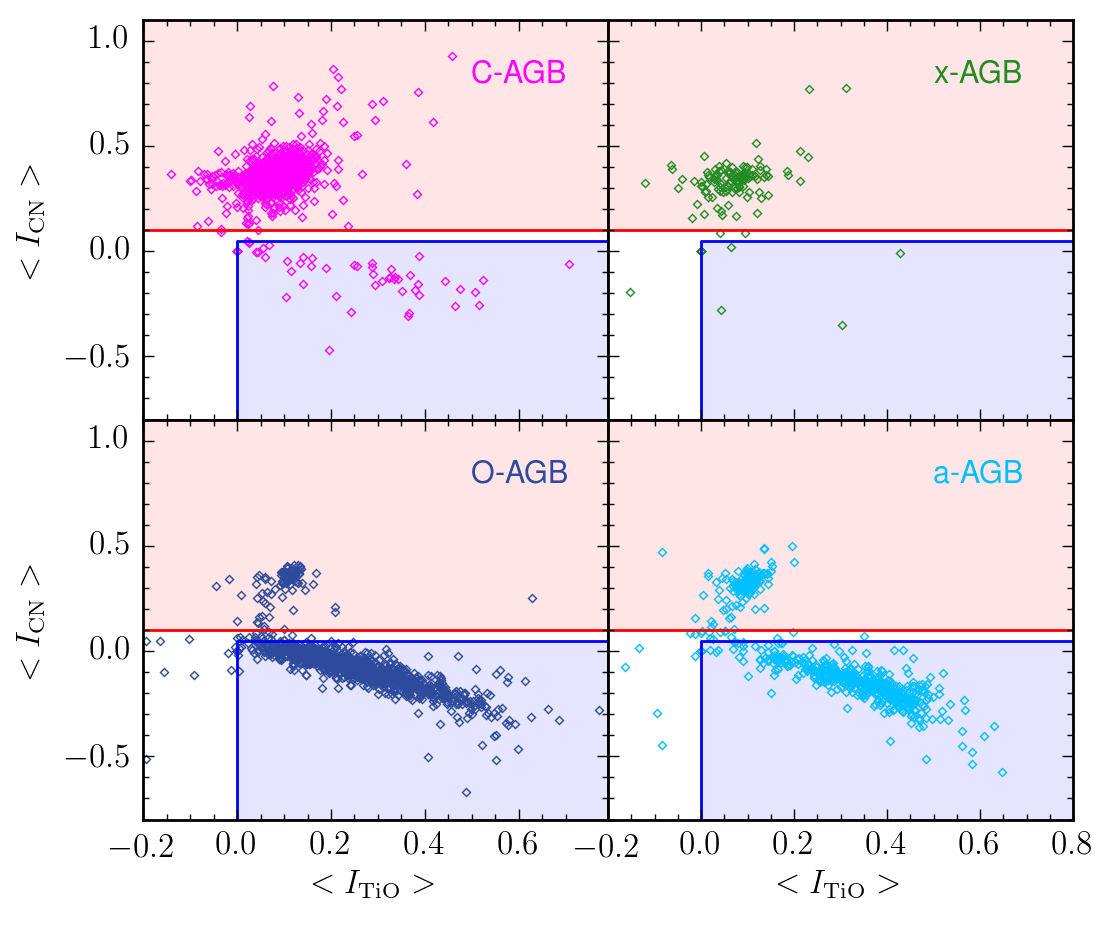}
\figcaption{ TIO vs. CN spectral indices for LMC stars compared to the
  photometric classifications from B11. Each subpanel includes sources
  that were photometrically classified as indicated by the label in
  the upper right. The blue and red shaded regions illustrate the
  spectral index classification employed here for O-rich and C-rich
  sources, respectively.  We classified stars in the white area by
  eye. This diagram demonstrates the cross-contamination between
    the photometric C-AGB and O-AGB groups (at the $\approx$4\%-9\%
    level), and shows that the a-AGB photometric class includes a
    sizeable fraction of both O-rich and C-rich sources. Also see
    Table~\ref{tab:class}. \label{fig:ind_lmc}}
\end{figure}

The Hydra-CTIO spectra (LMC AGB stars only) are classified by
computing band-strength indices covering TiO and CN to measure the
strength of these spectral features (Fig.~\ref{fig:knut};
Table~\ref{tab:index}). We calculate the index for each feature as
follows:

\begin{equation}
I = (F_{\rm cont.}^{\rm min} - F_{\rm band}^{\rm min})/F_{\rm cont.}^{\rm min},
\end{equation}

\noindent where $F_{\rm cont.}$ and $F_{\rm band}$ are the normalized
flux in the continuum and band ranges, respectively
(Table~\ref{tab:index}). The only S-type molecular features present in
the wavelength region covered by the CTIO spectra are due to LaO at
7385\AA\ and 7406\AA, but these features are weak and cannot be
reliably identified in this sample. We are thus limited to classifying
the LMC stars as either C-rich (${\rm C/O} > 1$) or O-rich (${\rm C/O}
< 1$).  We classify sources with $\langle I_{\rm CN} \rangle > 0.1$ as
C-rich and sources with $\langle I_{\rm CN} \rangle < 0.05$ and
$\langle I_{\rm TiO} \rangle > 0$ as O-rich.  Figure~\ref{fig:ind_lmc}
shows the classification of the LMC spectra compared to the
photometric classifications from B11 (Sect.~\ref{sec:photclass}). The
B11 photometric classifications agree with the spectral
classifications for 96\% of the C-AGB stars and 91\% of the O-AGB
stars (see Table~\ref{tab:class}).
 
There are 118 spectra that do not meet one of the spectral index
criteria described above, and we classify these by eye (i.e.,
  those in the white areas of Fig.~\ref{fig:ind_lmc} -- note that some
  fall outside of the plotted range).  We also visually inspect and,
if necessary, adjust the classification for the stars with a
spectroscopic index that disagrees with the photometric classification
(see Fig.~\ref{fig:ind_lmc}), which included 55 and 206
photometrically classified O-AGB and C-AGB stars, respectively.

The x-AGB stars are faint in the optical, resulting in noisy spectra
that make spectral index analysis difficult. We therefore check by
eye, and revise if necessary, the spectral classification of all x-AGB
stars. We are able to identify carbon or oxygen molecular features in
116 of the 258 x-AGB stars in the sample. All but four of these 116
are C-rich; the O-rich x-AGB stars include SSTISAGEMA
J045414.27$-$684413.9, SSTISAGEMA J051333.64$-$683644.4, SSTISAGEMA
J052710.19$-$693626.9 (WOH\,G339), and SSTISAGEMA
J044525.22$-$704648.3. The first of these is a known O-rich Mira
variable, with strong 10-\micron\ silicate emission \citep[alternate
  identifications: LI-LMC\,153, SP77\,30-6, and
  IRAS\,04544-6849;][]{Loup+1997,vanLoon+99,Dijkstra+2005}. WOH\,G339
is also known to show weak silicate emission \citep{Sloan+08}.

\begin{deluxetable}{lrr}
\tablewidth{0pc}
\tabletypesize{\normalsize}
\tablecolumns{3}
\tablecaption{Spectral Indices for LMC Spectra\label{tab:index}}

\tablehead{\colhead{Feature} &
\colhead{Band} &
\colhead{Continuum}
\\ 
&
\colhead{$\lambda$ (\AA)} & 
\colhead{$\lambda$ (\AA)}
} 

\startdata
TiO-1 & 7051---7057 & 7014---7051\\
TiO-2 & 7122---7128 & 7014---7051\\
TiO-3 & 7700---7750 & 7480---7530\\
CN-1 & 7250---7300 & 7215---7240\\
CN-2 & 7860---7920 & 7800---7845 

\enddata
\end{deluxetable}

\section{Analysis: Properties of a-AGB Stars}
\label{sec:analysis}

In this section, we summarize the observable properties of a-AGB stars.

\begin{figure}[h!]
\epsscale{1.1}
\includegraphics[width=\columnwidth]{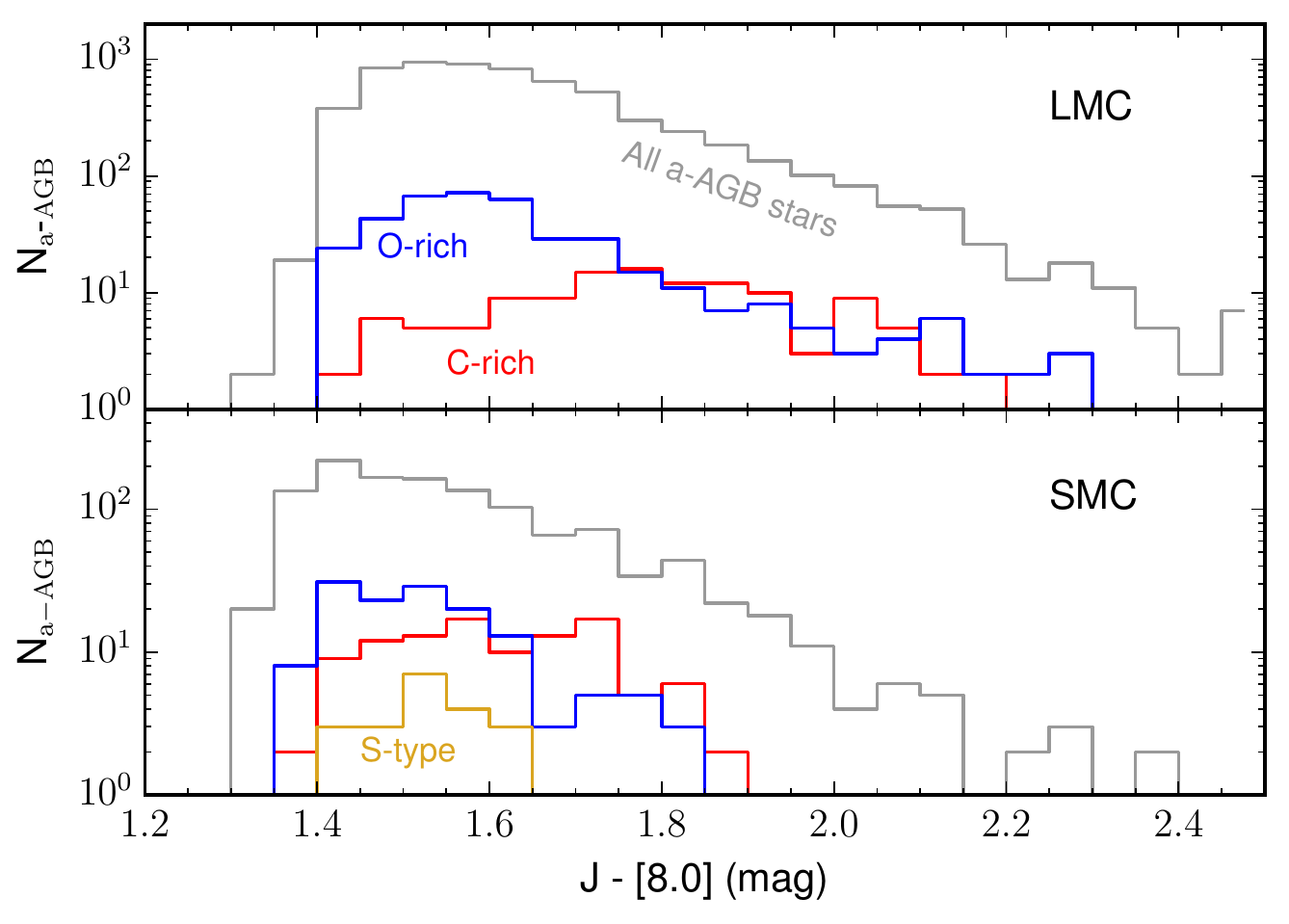}
\figcaption{ $J-[8]$ histogram of a-AGB stars in the LMC and
  SMC. Stars classified as O-rich or C-rich by their optical spectra
  are plotted in blue and red, respectively.  In the SMC, the blue
  line includes S-type and M-type stars, and S-type stars are also
  plotted separately in orange. S-type stars cannot be identified in
  the LMC sample due to limited wavelength coverage. In both galaxies,
  the bluest a-AGB stars tend to be O-rich. \label{fig:j8hist}}
\end{figure}

\begin{figure*}
\epsscale{1.1} 
    \hbox{\includegraphics[width=0.33\textwidth]{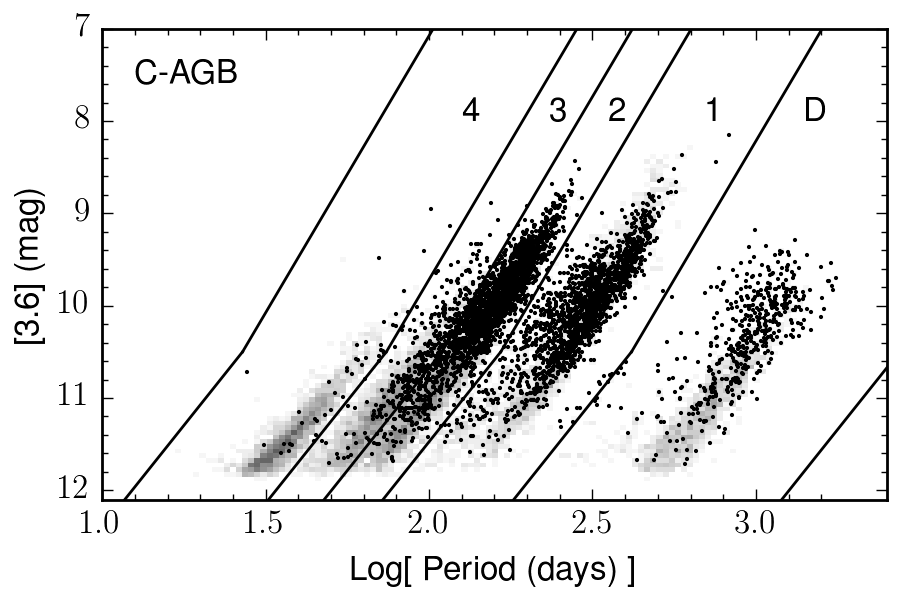}
    \includegraphics[width=0.33\textwidth]{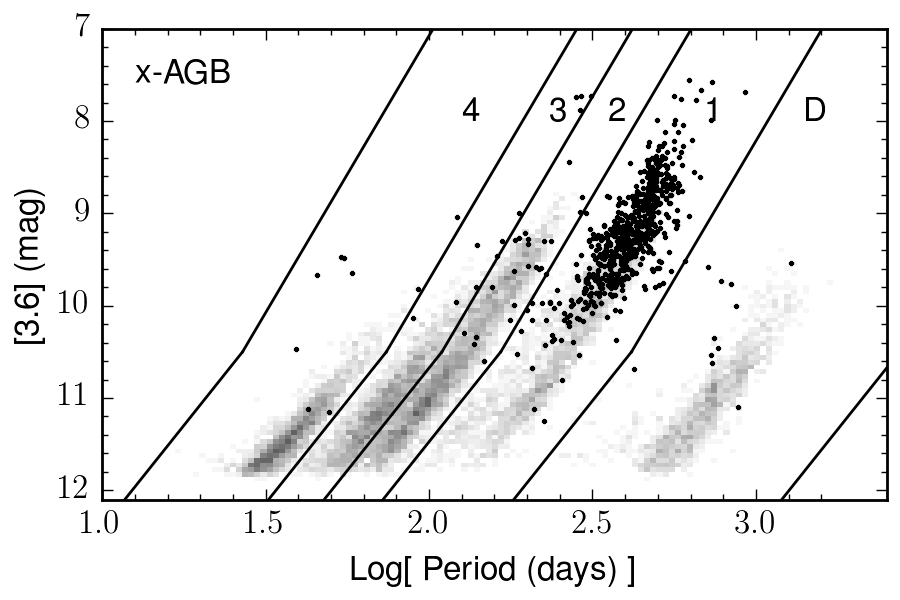}
\includegraphics[width=0.33\textwidth]{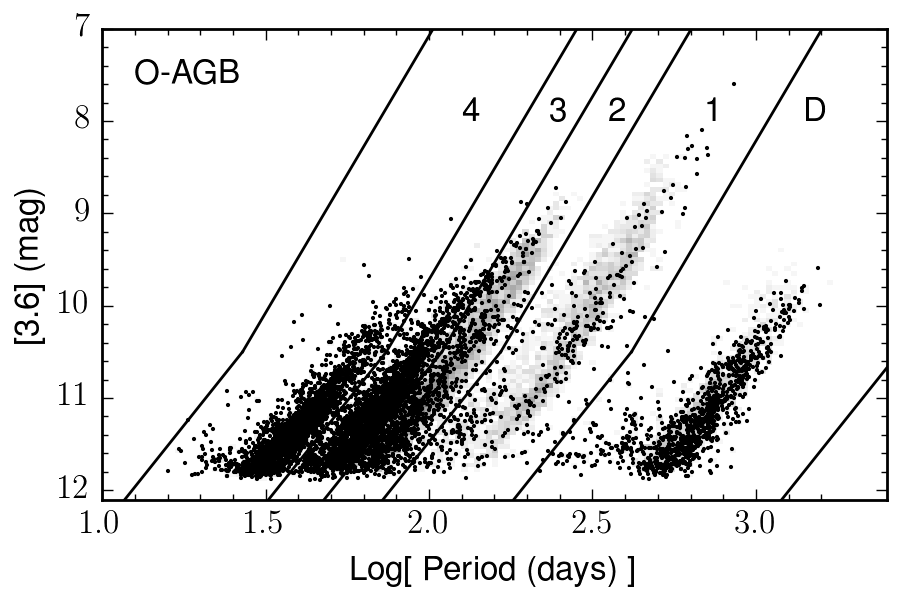}}
 \figcaption{3.6~\micron\ period-luminosity diagrams for the LMC
   (greyscale), separated by photometrically-classified spectral
   type (black dots).  For the O-AGB and C-AGB stars, we
   exclude sources that are spectrally classified as the other. The
   a-AGB stars are shown in
   Figure~\ref{fig:ao_pl}. \label{fig:agb_pl}}
\end{figure*}

\subsection{Chemistry}
\label{sec:chem}

Table~\ref{tab:class} summarizes the spectral classifications of
  the a-AGB stars. In the LMC, 76\% of the a-AGB spectra are O-rich,
and this fraction decreases to 45\% in the SMC (including M and S
stars).  In Figure~\ref{fig:j8hist}, we show the histogram of a-AGB
stars in $J-[8]$. In both galaxies, a-AGB stars are equally likely to
be C-rich or O-rich at red colors ($J-[8]>1.75$~mag in the LMC and
$>$1.55~mag in the SMC), and O-rich stars dominate at bluer colors,
though to a lesser degree in the SMC.

The high fraction of C-rich stars among the a-AGB sample indicates
that caution is necessary when using the standard $J-K_{\rm S}$ color
method to separate O-rich and C-rich stars.  Although the spectra
indicate that only about half of the a-AGB stars are O-rich in the
SMC, more than 96\% of them would classify as O-rich with the color
cuts suggested by \citet{Cioni+06b}. The exact placement of these cuts
also varies depending on the chosen foreground extinction correction
and the assumed metallicity of the galaxy, and slight shifts result in
potentially large uncertainties in the C/M ratio. By assuming that all
a-AGB stars are O-rich, B11 found C/M ratios of 0.56 in the SMC and
0.42 in the LMC. After correcting for the fraction of a-AGB stars
that are spectrally classified here as C-rich, we find revised C/M ratios
of 0.87 and 0.56, respectively. Relying on the $J-K_{\rm S}$ color is
therefore not ideal for separating C and M stars.

Between M and C stars lie the S-type stars; as an AGB star evolves,
carbon and {\it s}-process elements are dredged up to the surface.  As
C/O in the stellar envelope approaches unity, TiO bands in M stars
begin to disappear in favor of ZrO bands, which are prominent in
S-type AGB stars (though C/O can be near unity even in the absence of
ZrO bands). As more carbon is dredged up, free oxygen is tied up into
CO and carbon molecules dominate the stellar spectrum.  ZrO features
are prominent in two types of S stars: intrinsic and
extrinsic. Intrinsic S stars follow the evolution sequence just
mentioned. Extrinsic S stars are enriched in carbon and {\it
  s}-process elements through accretion from a binary companion that
has already finished its AGB evolution. Extrinsic S stars are less
evolved than intrinsic S stars, and are usually found on the RGB or
among early-AGB stars that have not yet begun to thermally pulse.  To
identify intrinsic S stars, studies rely on (1) the presence of
technetium, which will have decayed in the time since the mass
transfer episode that created the extrinsic S star
\citep[e.g.,][]{Jorissen2003}, (2) infrared colors that indicate the
presence of dust, which cannot form in less-evolved RGB and E-AGB
stars \citep[e.g.,][]{Groenewegen1993, Jorissen+1993,Otto+2011}, and
(3) high luminosities that are characteristic of thermally-pulsing AGB
stars \citep[e.g.,][]{GuandaliniBusso2008}.

\citet{Yang+06} identified Galactic intrinsic and extrinsic S-type AGB
stars based on their IR colors. B11 noted that the colors of the
a-AGB stars are consistent with the \citet{Yang+06} extrinsic S
stars, but without spectroscopy, we could not confirm or rule out the
presence of S stars among the a-AGB sample. Here, we find ZrO
features in the spectra of only 9\% of our SMC sample of a-AGB stars,
so an S-type nature cannot explain the red $J-[8]$ colors of the
entire a-AGB population. Moreover, despite their similar colors to
Galactic extrinsic S stars, the S stars detected here are likely to be
{\it intrinsic} because their luminosities (Section~\ref{sec:param})
and pulsation properties (Section~\ref{sec:pulse}) are consistent with
AGB stars undergoing thermal pulses. The overlap in color between the
a-AGB stars and the \citet{Yang+06} extrinsic sample suggests that IR
colors may not be a reliable way to separate intrinsic and extrinsic S
stars.

\begin{deluxetable}{rccccc}
\tablewidth{0pc}
\tabletypesize{\normalsize}
\tablecolumns{6}
\tablecaption{3.6~\micron\ $P$-$L$ Relationship Statistics\label{tab:PL}}

\tablehead{\colhead{AGB} &
\colhead{Seq. 4} &
\colhead{Seq. 3} &
\colhead{Seq. 2} &
\colhead{Seq. 1} &
\colhead{Seq. D}
\\
\colhead{class}&
\colhead{N\,(\%)} &
\colhead{N\,(\%)} &
\colhead{N\,(\%)} &
\colhead{N\,(\%)} &
\colhead{N\,(\%)} 
} 

\startdata
\multicolumn{6}{c}{-- Photometric Classifications --}\\
\multicolumn{6}{c}{LMC}\\
a   &  213\,(4)  &  448\,(9)  & 2135\,(43) & 1140\,(23) & 1003\,(20) \\
O   & 2756\,(37) & 2248\,(30) & 1066\,(14) & 373\,(5)   & 1002\,(13) \\
C   &   41\,(1)  &  297\,(7)  & 2096\,(46) & 1634\,(36) &  470\,(10) \\
x   &     9\,(1) &   10\,(1)  &   23\,(3)  & 656\,(92)  &   12\,(2)  \\
\multicolumn{6}{c}{SMC}\\
a   &  56\,(6)  & 118\,(12) & 317\,(32) & 160\,(16) & 355\,(35) \\
O   & 566\,(32) & 443\,(25) & 224\,(13) & 103\,(6)  & 448\,(25) \\
C   &   7\,(1) &  53\,(4) & 588\,(40) & 603\,(41) & 208\,(14) \\
x   &   0\,(0) &   2\,(1) &   4\,(2) & 219\,(96) &   2\,(1)\\
\hline\\
\multicolumn{6}{c}{-- Spectral Classifications; a-AGB only --}\\
\multicolumn{6}{c}{LMC}\\
O & 6\,(3) & 15\,(6)& 99\,(43)& 63\,(27) & 48\,(21) \\
C & 7\,(9) & 19\,(25) & 19\,(25) & 11\,(15) & 19\,(25) \\

\multicolumn{6}{c}{SMC}\\
O & 5\,(6) & 5\,(6) & 29\,(32) & 9\,(10) & 42\,(47) \\
C & 14\,(13) & 28\,(26) & 24\,(22) & 13\,(12) & 29\,(27)\\
S & 0\,(0) & 1\,(6) & 12\,(57) & 3\,(14) & 5\,(24) 
\enddata
\tablecomments{\ Position of AGB stars on the 3.6~\micron\ {\it P-L}
  diagram. The percentage of the total population of a particular
  subtype is shown in parentheses. For example, 213 of 4939 a-AGB
  stars (4\%) are found on Sequence 4 in the LMC (see Table~\ref{tab:class}). For an independent
  accounting of the pulsation sequences, see \citet{Spano+2011}.}

\end{deluxetable}

\begin{figure}
\epsscale{1.1}
\includegraphics[width=\columnwidth]{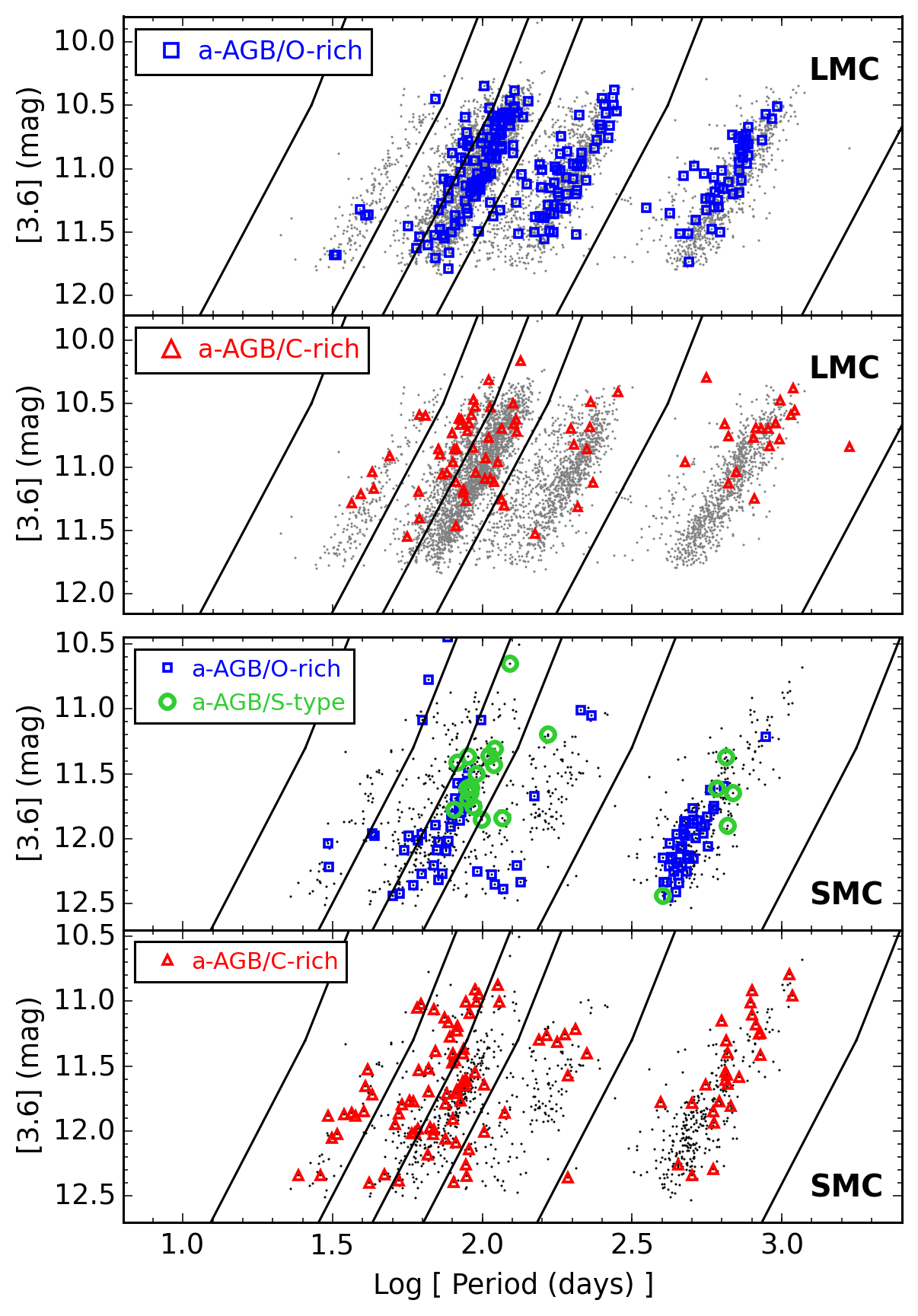}
\figcaption{3.6~\micron\ period-luminosity diagrams for a-AGB
  stars. Grey dots show all a-AGB stars, and colored symbols mark
  those with spectral classifications. Note the change in y-axis from
  Figure~\ref{fig:agb_pl}, and the y-axis change between the LMC and
  SMC. \label{fig:ao_pl}}
\end{figure}

\subsection{Pulsation Properties}
\label{sec:pulse}

Figure~\ref{fig:agb_pl} shows the LMC 3.6~\micron\ $P$-$L$ diagram for
each type of (photometrically-classified) AGB star; the proportions of
stars on each sequence in the SMC are similar. Table~\ref{tab:PL}
shows that the a-AGB stars preferentially occupy Sequences~1, 2, and
D (we discuss Sequence~D stars in Section~\ref{sec:seqD}).  Most
a-AGB stars are classified by OGLE as semi-regular variables (SRVs),
which have pulsation properties similar to Miras, but with less
regular light curves and smaller $V$-band amplitudes.

In both galaxies, the majority of the fundamental-mode pulsators
(Sequence~1; those at the end of the TP-AGB phase) are photometrically
classified as C-rich (either C-AGB or x-AGB). Only 9--10\% of them are
photometrically classified as O-rich, and 15--30\% are a-AGB
stars. Almost 90\% of the O-AGB stars have shorter periods
(Sequences~3 and 4), indicating that the O-AGB class as defined in B11
comprises mostly stars at the beginning of the TP-AGB phase.

In Figure~\ref{fig:ao_pl}, we show the 3.6~\micron\ $P$-$L$
relationship for a-AGB stars classified via their optical spectra.
In both galaxies, the O-rich a-AGB stars (with TiO or ZrO features)
lie on sequences 2 and D. In the LMC, they also occupy Sequence 1.
Pulsation models \citep{Wood2015} show that stars move to longer
periods as their luminosities increase, suggesting that many high-L
a-AGB stars on Sequence~2 will evolve into C stars as they cross to
Sequence~1 and that most of the O-rich a-AGB stars on Sequence~1 will remain
O-rich. The absence of a-AGB stars on Sequence~1 in the SMC reflects
the efficiency of this transition in metal-poor environments.

We have identified 24 S-type stars in the SMC, and 22 of them are
classified as AGB stars by the OGLE-III survey (the remaining 2 fall
just outside the OGLE-III spatial coverage). The OGLE light curves thus
provide additional evidence that the {\it s}-process elements evident
in their spectra are from the dredge-up of carbon during a thermal
pulse (intrinsic S-type), and not due to mass transfer (extrinsic
S-type). The clustering of these intrinsic S stars on the
3.6-\micron\ $P$-$L$ diagram is an intriguing clue about the dredge-up
process.  However, the SMC spectral sample does not include stars from
the C-AGB and O-AGB photometric categories, so the frequency of S-type
stars among those groups remains unknown, as does the position of any
such stars on the $P$-$L$ diagram.

\subsection{Fundamental Parameters: L, $T_{\rm eff}$, M, and [Fe/H]}
\label{sec:param}

\begin{figure}
  \includegraphics[width=\columnwidth]{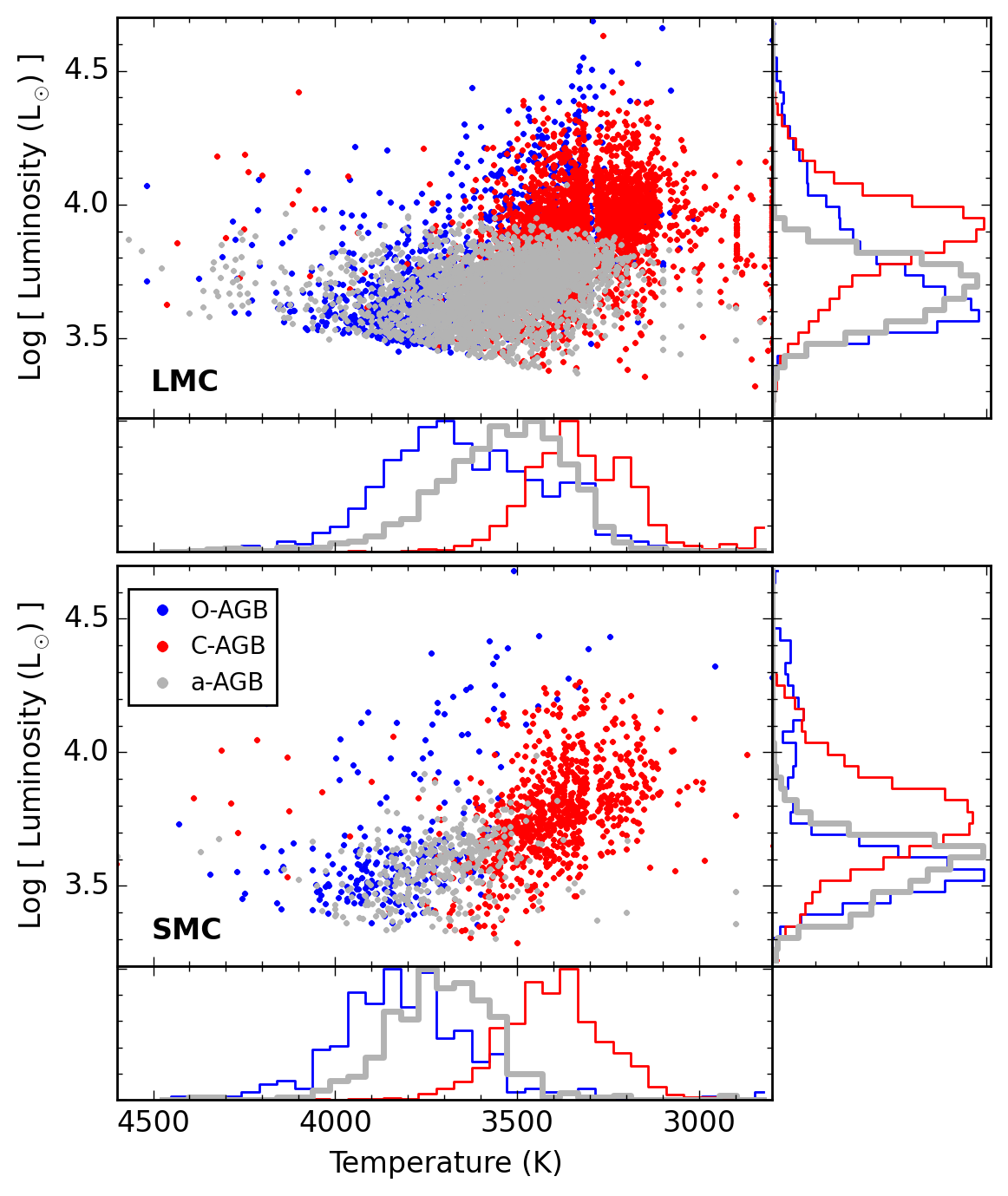}
  \figcaption{Hertzprung-Russell diagram for stars on pulsation
    Sequences 1 and 2. The x-AGB stars are excluded because
    circumstellar extinction results in unreliable SED fits. The
    a-AGB stars are intermediate to the O-AGB and C-AGB stars in both
    $T_{\rm eff}$ and $L$.
\label{fig:HR}}
\end{figure}

To estimate the luminosities ($L$) and effective temperatures ($T_{\rm
  eff}$), we fit the spectral energy distributions (SEDs) of the AGB
stars using the photometry listed in Table~\ref{tab:phot} and
following the procedure outlined by \citet{McDonald+09,McDonald+2012}.
Each SED was compared to a grid of {\sc bt-settl} stellar atmosphere
models \citep{Allard+2011} using $\chi^2$ minimization and scaled in
flux to derive a bolometric luminosity. We assume ${\rm [Fe/H]} = -0.3
  (-0.5)$, ${\rm [\alpha/Fe]} = +0.3 (+0.3)$, $d=51$~kpc (61~kpc), and
    $E(B-V) = 0.085$~mag (0.070~mag) for the LMC (SMC).

Uncertainties are large for stars with significant circumstellar dust
since this method compares SEDs to dustless photospheres. In practice,
we find that SED fits with temperatures less than 2800~K, encompassing
most of the x-AGB stars, are unreliable. We therefore exclude most
x-AGB stars when discussing effective temperatures, luminosities, and
pulsation masses. 

We show the Hertzprung-Russell diagram in Figure~\ref{fig:HR}. To
compare stars in a similar evolutionary phase, we include only stars
on pulsation Sequences~1 and 2. In both galaxies, the a-AGB stars
have temperatures and luminosities intermediate to the O-AGB and C-AGB
stars. This suggests that a-AGB stars are those that are near a
transition from O- to C-rich.  The temperatures and luminosities of
the a-AGB stars here are similar to those determined for the C-rich
and S-type stars in the {\it Hipparcos} catalog using the same method
\citep{McDonald+2012}.

To derive the stellar masses, we rely on the period-mass-radius
relationship described by \citet{VassiliadisWood93}. This relationship
is valid only for stars that lie on pulsation Sequence 1 and are
therefore pulsating in the fundamental mode:

\begin{equation}
\label{eq:pmr}
\log P ({\rm d}) = \\-2.07 + 1.94 \log R/R_\odot - 0.9 \log M_{\rm
  p}/M_\odot,
\end{equation}

\noindent with $R$ derived using $L \sim R^2 T_{\rm eff}^4$. The mass
derived from this relationship is the pulsation mass ($M_{\rm p}$),
which reflects the current mass of the star, so small masses could
either indicate a low-mass star that has undergone little mass loss or
a high-mass star that has undergone substantial mass loss.  

\begin{figure}
  \includegraphics[width=\columnwidth]{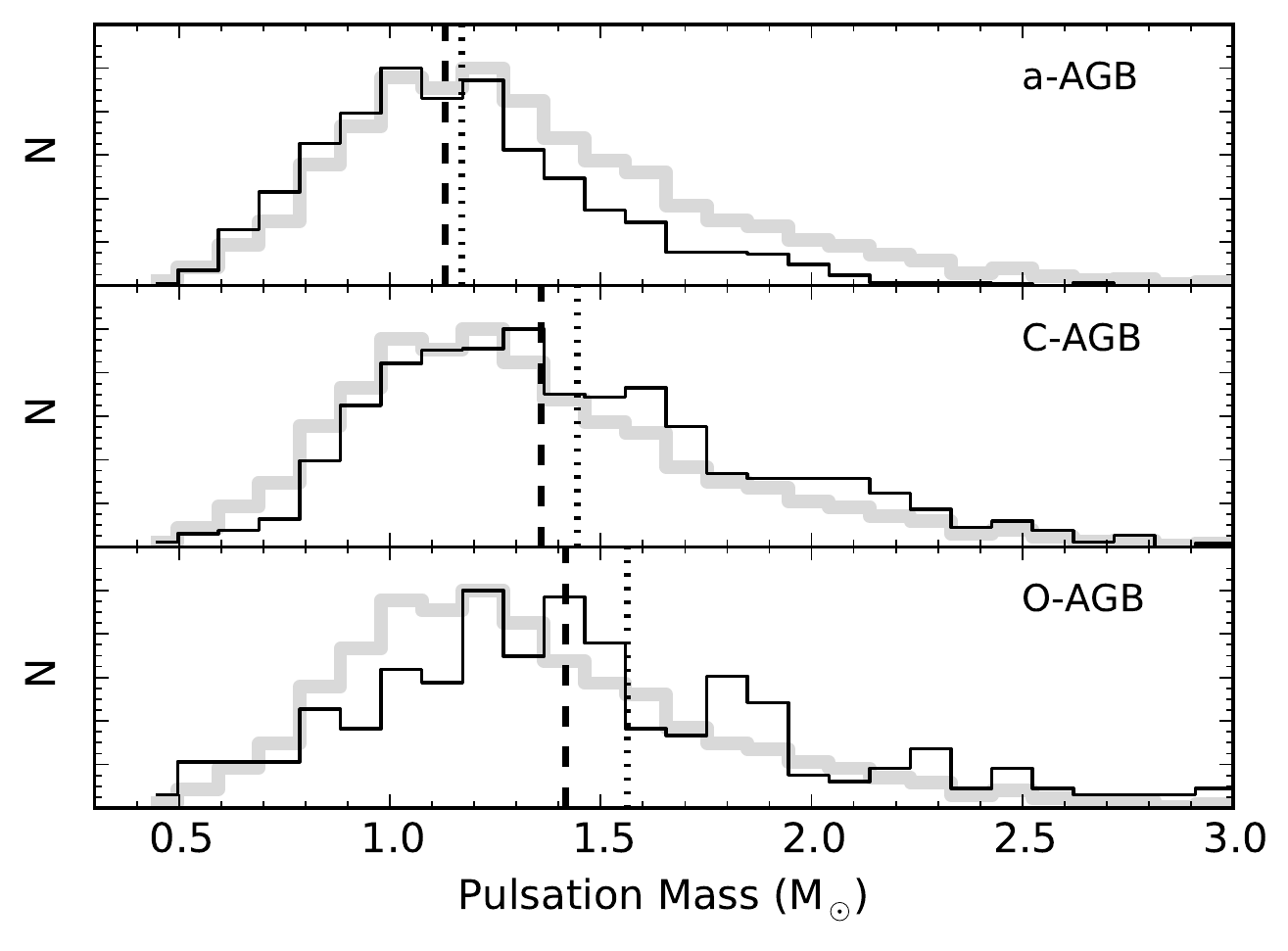}
  \figcaption{Distribution in pulsation mass for stars on Sequence~1
    in the LMC. Stars included here are photometrically
    classified. The thick gray line shows the sum of all three
    components, the dotted line marks the mean, and the dashed line
    marks the median.
\label{fig:mass}}

\end{figure}

Figure~\ref{fig:mass} shows the distribution in pulsation mass for the
fundamental-mode pulsators, divided by stellar type.  Most stars
(95\%) have pulsation masses of 0.5--2.5~$M_\odot$, in agreement with
\citet{Wood2015}, and the a-AGB stars show the lowest median
pulsation masses. Since the a-AGB mass-loss rates are not
particularly high (see Section~\ref{sec:mlr}), we can infer that the
a-AGB stars have the lowest {\it initial} masses among the stars
pulsating in the fundamental mode.

Metallicities were derived using the Ca triplet by \citet{Olsen+11},
including 36 a-AGB stars and 268 O-AGB stars in the LMC. Abundances
were not reliably measured for any C-AGB or x-AGB stars. The median
metallicity of the O-AGB population is ${\rm [Fe/H]} = -0.82$~dex,
with dispersion (median absolute deviation) of 0.38~dex and a median
uncertainty of 0.17~dex.  The a-AGB population is more metal-poor,
with a median ${\rm [Fe/H]} = -1.05$~dex, dispersion 0.36~dex, and
median uncertainty 0.19~dex. The lower metallicities suggest that the
a-AGB population is older, having formed earlier in the LMC's
chemical evolution history. This finding is consistent with the lower
masses inferred from the pulsation. Alternatively, the a-AGB stars
may have been accreted from the SMC or formed from gas accreted from
the SMC.

\subsection{Mass Loss and Dust Production}
\label{sec:mlr}

Fourteen of the LMC a-AGB stars were observed with {\it Spitzer's}
infrared spectrograph (IRS) from 5--16~\micron\ as part of the
SAGE-Spec program \citep{Kemper+10}. The equivalent program for the
SMC does not include any a-AGB stars \citep{Ruffle+2015}. Of the 14
LMC stars, four are classified by \citet{Woods+2011} as Stars (their
AGB nature is unconfirmed) and the remaining ten as O-rich AGB stars,
nine of which show strong silicate emission at 10~\micron\ (see
Section~\ref{sec:cause}).  With only 14 examples, we cannot draw any
broad conclusions about the IR spectral features in a-AGB stars, but
the presence of a silicate feature in these examples demonstrates that
a-AGB stars are capable of significant dust production.  

\begin{figure}
  \includegraphics[width=\columnwidth]{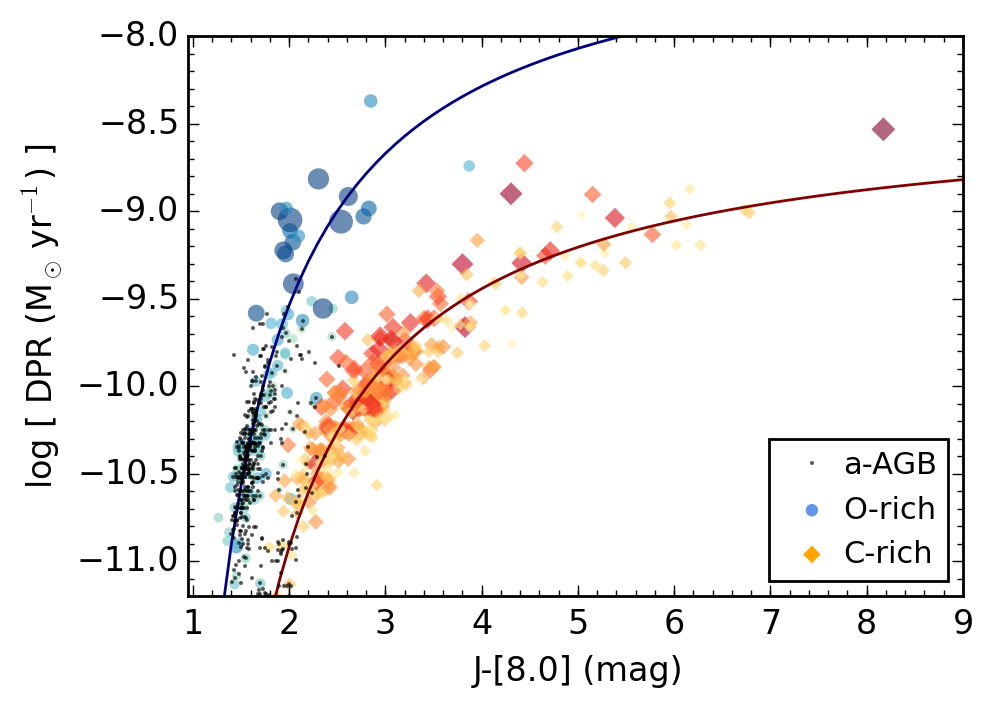}
  \figcaption{The color-DPR relationship for LMC stars on
    Sequence~1. The blue and orange points were classified by their
    spectra, and the black dots are the photometrically-identified
    a-AGB stars (all Sequences).  In the LMC, most of the a-AGB
    stars follow the low-mass end of the O-rich
    branch. \label{fig:dprsum}}
\end{figure}

\citet{Riebel+2012} computed the mass-loss and dust-production rates
(DPRs) for the LMC AGB stars by fitting the full optical to mid-IR
SEDs to the Grid of AGB and RSG ModelS
\citep[GRAMS;][]{Sargent+2011,Srinivasan+2011}. The GRAMS grid
includes the COMARCS stellar photospheres for C stars
\citep{Gautschy+04} and PHOENIX models for O-rich stars
\citep{Hauschildt+99}. Silicate dust or a mixture of amorphous
carbon$+$SiC were added to create a grid that reproduced the full
luminosity and infrared color distribution of the LMC AGB
population. We note that luminosities derived with GRAMS using a
simple trapezoidal integration are systematically lower than those
derived by the SED fitting (Section~\ref{sec:param}) by about 16\% for
O-rich stars and 18\% for C-rich stars.

For a subset of sources, the GRAMS classification from
\citet{Riebel+2012} disagrees with our spectral classification. This
includes 29 of the spectrally-confirmed O-rich sources that were
classified by GRAMS as C-rich, and 241 of the spectrally-confirmed
C-rich sources that were likewise classified as O-rich by GRAMS.
GRAMS provides a best-fit O-rich and C-rich model for each star, and
the GRAMS classification is assigned depending on which provides a
smaller $\chi^2$, so many misclassifications are the result of only a
very slight preference for one fit over the other. Here, we assign the
best-fit C- or O-rich GRAMS model that matches the spectral
classification.

The GRAMS grid assumes a wind expansion velocity of $v_{\rm
  exp}=10$~km\,s$^{-1}$, a spherical dust shell geometry, and a
constant DPR throughout the AGB lifetime. The DPR scales linearly with
$v_{\rm exp}$, which itself scales with luminosity and the gas-to-dust
ratio ($\psi$) as $v_{\rm exp} \propto L^{0.25} \psi^{-0.5}$. To scale
the GRAMS DPRs, we use different gas-to-dust ratios depending on the
stellar chemistry. For O-rich stars, we assume that $\psi$ scales with
metallicity, so $\psi_{\rm LMC}^{\rm O\mbox{-}rich} = 400$ for $Z_{\rm
  LMC} = 0.5\,Z_\odot$ and $\psi_\odot = 200$ (here we use the
typically assumed LMC metallicity).  For a-AGB stars, we scale $\psi$
according to the difference in metallicities between O-AGB and a-AGB
stars discussed in Section~\ref{sec:param}, so $\psi_{\rm LMC}^{\rm
  a\mbox{-}AGB} = 680$. Since C stars make their own carbon, we
expect their gas-to-dust ratios to be near solar, so we use $\psi_{\rm
  LMC}^{\rm C\mbox{-}rich} = 200$.

Estimated DPRs from SED fitting are highly sensitive to the choice of
dust optical constants and dust compositions, which may differ at
these metallicities \citep[e.g.,][]{McDonald+2010,Jones+2014}. We
therefore note that while the {\it relative} GRAMS-derived DPRs for
different stellar types are valid, {\it absolute} DPRs are inaccurate
by a factor of $\sim$4
\citep{Groenewegen+09,Srinivasan+2011,Sargent+2011,McDonald+2011b}.

The DPRs indicate that the a-AGB stars are moderately strong dust
producers \citep[also see][]{Boyer+2012}. The median DPR of the a-AGB
population is $4 \times 10^{-11}~M_\odot\,{\rm yr}^{-1}$ (dispersion
$2 \times 10^{-11}~M_\odot\,{\rm yr}^{-1}$).  This is an order of
magnitude higher than the median DPR of the O-AGB population, which
includes many non-dusty stars, and a few bright very dusty stars.  The
C-AGB and x-AGB populations are very dusty, with median DPRs that are
$\approx$2$\times$ and 2 orders of magnitude higher than the a-AGB
median DPR, respectively.

Figure~\ref{fig:dprsum} shows how the $J-[8.0]$ color varies with the
DPR; colors represent the spectral classification and symbol sizes
represent the pulsation mass (only Sequence~1 stars are plotted, so
the low-DPR O-AGB population, which resides on Sequences~3 and 4 are
not included). The relationship between $J-[8]$ and DPR is bimodal,
with O-rich stars reaching higher DPRs at bluer colors than C stars
and a clean separation between the two spectral types. O-rich stars
increase in mass as the DPR increases, but the masses of C-rich stars
is mixed along the entire range of DPRs. Fits to each branch shown in
Figure~\ref{fig:dprsum} are of the form:

\begin{equation}
\label{eq:color}
\log \dot{D} = \frac{x}{(J-[8.0])+y} + z
\end{equation}

\noindent for stars with $J-[8.0]>1$~mag. For C-rich stars, $x=-3.60$,
$y=-0.58$, and $z=-8.39$. For O-rich stars, $x=-3.54$, $y=-0.42$, and
$z=-7.30$, though it is unclear whether this relationship applies to
O-rich stars with $J-[8.0]>3$~mag, since we have few examples
of O-rich stars at those extreme colors.

Overplotted as dots in Figure~\ref{fig:dprsum} is the entire
population of a-AGB stars, including all pulsation sequences. The
bulk of the a-AGB stars smoothly continue the O-rich branch to low
DPRs and masses, indicating that the a-AGB stars are a subset of the
general (dust-producing) O-rich AGB population in the LMC.

\section{Discussion}
\label{sec:disc}

The properties of the a-AGB stars presented in the previous section
all suggest that they are low-mass, very evolved dusty AGB stars.  The
broad spatial distributions of the a-AGB stars compared to other AGB
stars in the SMC presented in B11 also suggests that a-AGB stars
originate from an older stellar population. The star-formation
histories of the Magellanic Clouds have been computed globally by
\citet{Harris+04,HarrisZaritsky2009}, and in smaller fields by
\citet{Rubele+2012,Rubele+2015} and \citet{Weisz+2013}. While there is
strong variation in the star-formation histories regionally, global
histories show a peak near $t_{\rm age} = 2$--$3$~Gyr, which
corresponds to an initial stellar mass of 1--1.3~$M_\odot$.  The
a-AGB stars likely originated from this star-formation event.  This
mass range encompasses the lower-mass limit for stars that will become
C-rich during their evolution \citep[e.g.,][]{Marigo+13}.

In this section, we explore additional evidence supporting this
low-mass hypothesis, compare the characteristics of the a-AGB stars
to their more massive counterparts, and discuss the opportunities that
arise with a simple photometric technique for distinguishing
low-mass TP-AGB stars. We note that the O-AGB population, as defined
in Section~\ref{sec:photclass} can be separated into more evolved
(pulsation Sequences 1 and 2), and less-evolved (pulsation Sequences 3
and 4) TP-AGB stars.  The following sections exclude the less evolved
O-AGB stars, except where explicitly stated.

\subsection{Dust Production and Pulsation}
\label{sec:dustpulse}

\begin{figure}
  \includegraphics[width=\columnwidth]{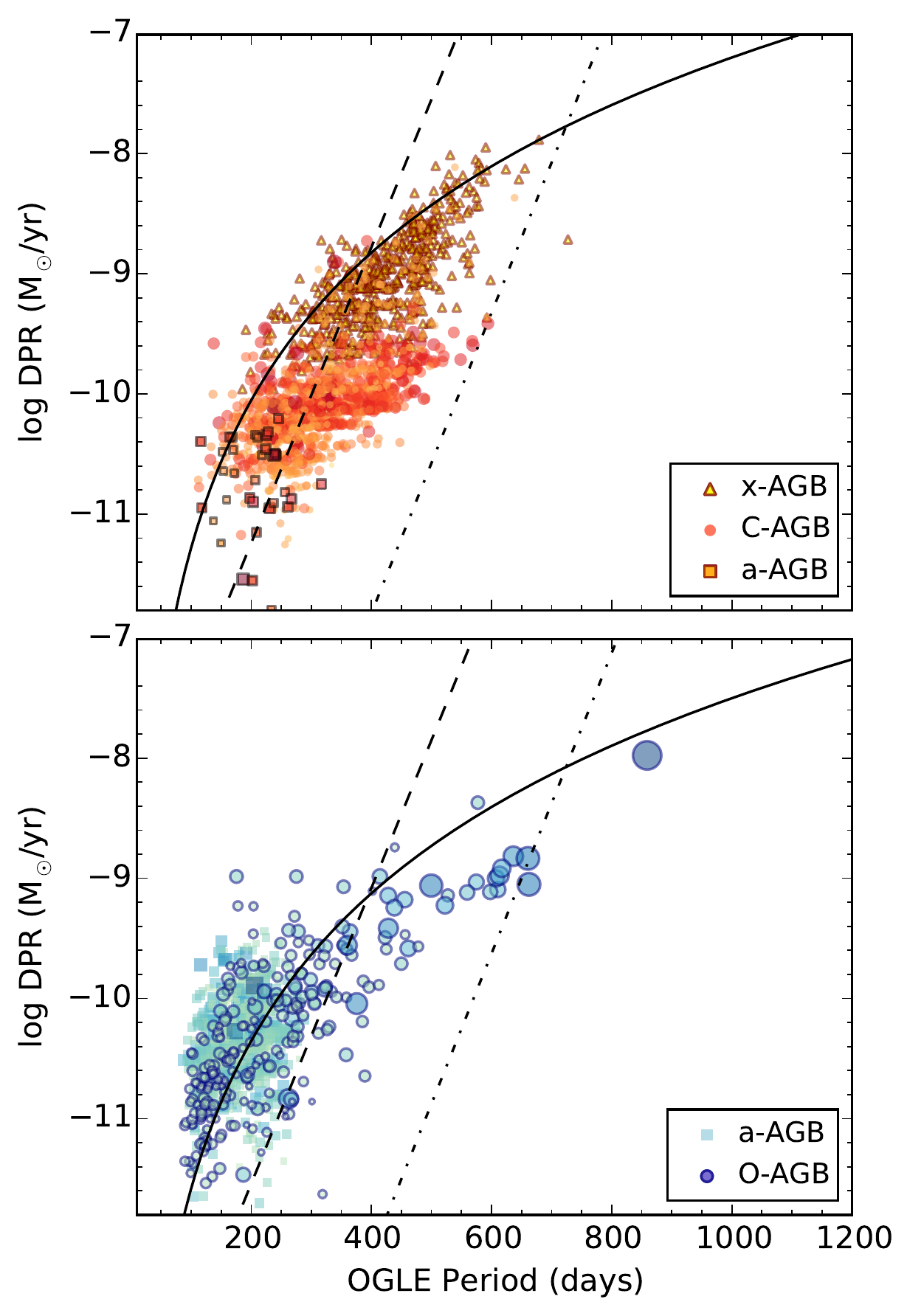}
  \figcaption{The relationship between pulsation period and mass loss
    for AGB stars pulsating in the fundamental mode (Sequence 1) in
    the LMC. The upper panel shows C-rich stars and the lower panel
    shows O-rich stars. Large symbols/darker colors indicate larger
    pulsation masses for C-AGB, O-AGB, and a-AGB stars; x-AGB stars
    are plotted with the same colors/sizes since their masses could
    not be reliable estimated.  In both panels, we include stars
    classified via photometry, and use the optical spectra to
    eliminate interlopers. The straight lines are from
    \citet{VassiliadisWood93} for stars with pulsation masses
    $<2.5~M_\odot$ (dashed line) and $=5~M_\odot$ (dot-dashed line),
    derived by fitting Galactic C-rich and O-rich Miras. The curved
    thin line is from \citet{Groenewegen+98}, derived from Galactic
    C-rich Miras and scaled down to account for the difference in
    optical constants \citep[see][]{Srinivasan+2011}. 
\label{fig:mod}}
\end{figure}

Several works point to pulsation as one of the drivers of AGB dust
production and mass loss, and we now explore the influence of
pulsation on the dust-production rates of a-AGB stars. The influence
of pulsation is clear: the median DPR increases by an order of
magnitude from Sequences~4 to 3, and again from Sequences~3 to 2, only
surpassing $10^{-11}~M_\odot\,{\rm yr}^{-1}$ on Sequences~2, 1, and D,
where the a-AGB stars reside.

\citet[][VW93]{VassiliadisWood93} found that for Galactic C-rich and
O-rich AGB stars pulsating in the fundamental mode, the mass-loss rate
increases exponentially with period as:

\begin{equation}
\label{eq:vw1}
\log \dot{M} (M_\odot\,{\rm yr}^{-1})= -11.4 + 0.0123 P ({\rm days}) 
\end{equation}

\noindent until $P=500$~days, when the star enters the superwind phase
and the mass-loss rate reaches a maximum near $10^{-4.2}~M_\odot\,{\rm
  yr}^{-1}$. For stars more massive than 2.5~$M_\odot$, their models
suggest that this relation changes to

\begin{multline}
\label{eq:vw2}
\log \dot{M} = -11.4 + 0.0125 [P - 100(M/M_\odot -2.5)].
\end{multline}

\noindent The VW93 parameterization is shown for the LMC in
Figure~\ref{fig:mod} along with the scaled dust-production rates from
\citet{Riebel+2012} and the periods from OGLE, using the spectral
classifications to exclude stars that are photometrically
misclassified. We convert the gas mass-loss rates from VW93 to DPRs
using a solar gas-to-dust ratio ($\psi=200$) for C stars and scaling
the ratio with metallicity for O-rich stars
($\psi=400$). Equation~\ref{eq:vw1} provides a reasonably good
description of the behavior of C-rich stars.  The
predicted trend for more massive stars to move to longer periods is
evident in Figure~\ref{fig:mod} for the O-rich stars.

\citet{Groenewegen+98} \citep[and later][]{Groenewegen+09} fit the
spectral energy distributions of several Galactic C-rich Miras to
revise the VW93 $\dot{M}$-$P$ relationship. They find

\begin{equation}
\label{eq:g09}
\log \dot{M} = (4.08 \pm 0.41) \log P - (16.54 \pm 1.10),
\end{equation}

\noindent which we show in Figure~\ref{fig:mod}, scaled down by a
factor of 4 to account for differences in optical constants as
described by \citet{Srinivasan+2011} and by the gas-to-dust ratio
($\psi_{\rm C}=200$; $\psi_{\rm O}=400$). The LMC AGB stars do not
deviate strongly from the G98 Galactic $\dot{M}$-$P$ relationship, despite
their lower metallicities ($Z_{\rm LMC} \sim 0.5 Z_\odot$).

The LMC a-AGB stars on pulsation Sequence~1 (and thus included in
Fig.~\ref{fig:mod}) are primarily O-rich. Their position to the left
of the dashed line from VW93 indicates masses $<$2.5~$M_\odot$ (the
O-AGB stars within the a-AGB locus could easily also be classified as
a-AGB stars with a slight shift in the color classifications -- see
Section~\ref{sec:aoclass}).  Unlike the more massive O-rich stars and
the x-AGB/C-AGB stars, the a-AGB DPRs are not strongly dependent on
the pulsation period: their DPRs span two orders of magnitude over the
full a-AGB period distribution. The DPRs of the x-AGB and massive
O-AGB stars, on the other hand, show a strong positive correlation
with both pulsation periods and amplitudes. These a-AGB star
characteristics suggest that once a low-mass star is pulsating in the
fundamental mode, the pulsation plays only a small role in its dust
production, at least compared to their more massive counterparts.

\begin{figure}
\vbox{
  \includegraphics[width=\columnwidth]{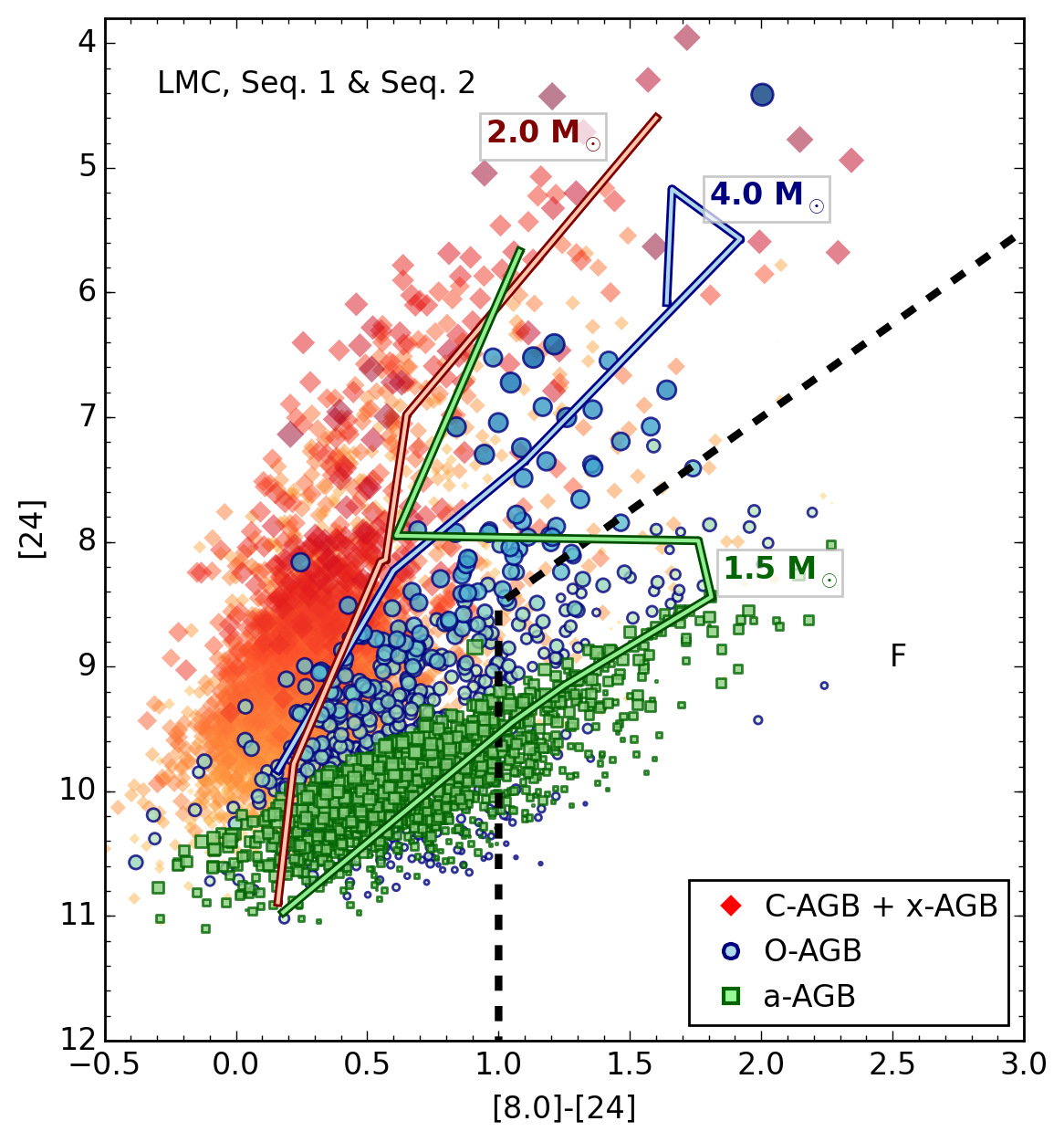}
  \includegraphics[width=\columnwidth]{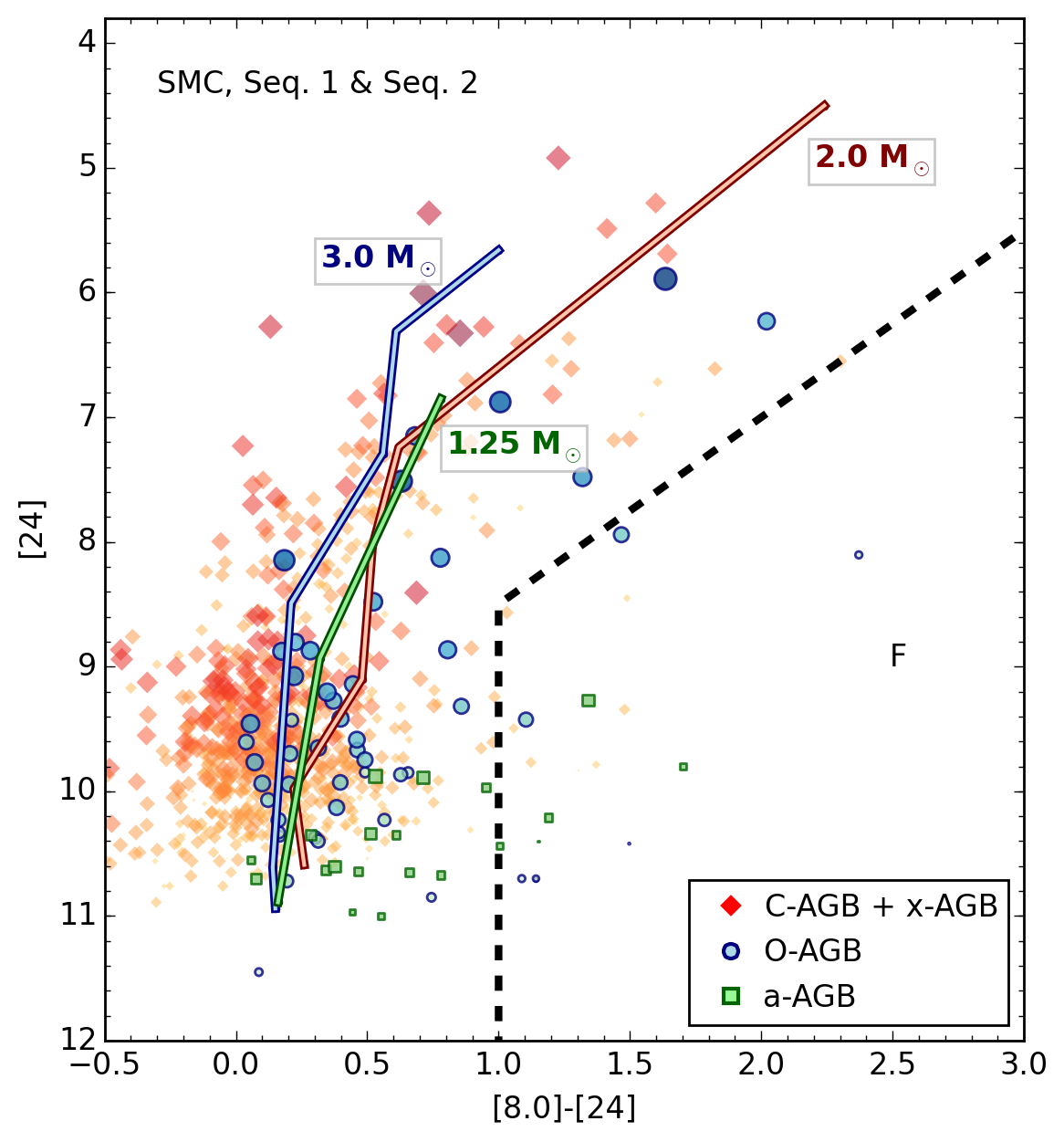}
} \figcaption{24~\micron\ CMD for AGB stars on Sequences~1 and
  2. Larger/darker symbols indicate a higher bolometric
  luminosity. Plotted stars are classified photometrically, using the
  available spectra to minimize misclassifications. Evolutionary
  tracks from \citet{DellAgli+2015} at $Z=0.004$ (SMC) and
  $0.008$ (LMC) are overplotted for the indicated initial
  stellar masses. The LMC a-AGB stars correspond to
  $\lesssim$1.5~$M_\odot$ tracks, while in the SMC, they must have
  initial masses $\lesssim$1.25~$M_\odot$, and/or rapidly transition
  to a C-rich atmosphere. The region to the right of the dashed line
  is occupied by the ``F'' finger, noted by \citet{DellAgli+2015} and
  encompasses the dustiest a-AGB stars.
\label{fig:dell1}}
\end{figure}

\subsection{Comparison to Stellar Evolution Models}

\citet{DellAgli+2015} (hereafter D15) computed theoretical
evolutionary tracks of AGB stars, including dust formation. They
found that the masses of AGB stars can be inferred by their positions
in a {\it Spitzer} $[8.0]-[24]$ vs. $[24]$ CMD. In the upper panel of
Figure~\ref{fig:dell1}, we show the CMD of the LMC for Sequence~1 and
2 stars. As stars increase in bolometric luminosity
(Section~\ref{sec:param}), the size of the symbol increases and the
color darkens. The most luminous (and by extension, the most massive)
stars are found at the bright/red end of the CMD, following the
predictions of D15.

The a-AGB stars occupy a finger at faint 24-\micron\ magnitudes, with
a broad range in $[8.0]-[24]$ color.  In D15, this region of the CMD
is populated by stars with initial masses $\lesssim$1.5~$M_\odot$,
prior to transitioning from O-rich to C-rich, which they designate
``F'' stars (dashed line in Fig.~\ref{fig:dell1}a). Their models
indicate that ``F'' stars evolve along this finger to brighter
24~\micron\ magnitudes before shifting horizontally to the C star
sequence when dredge-up causes C/O to exceed unity at the very end of
the AGB phase. Stars with lower masses, and thus insufficient dredge
up for ${\rm C/O}>1$, would remain on this finger. Indeed, dusty AGB
stars in Globular Clusters ($M_{\rm i} \lesssim 1~M_\odot$) also
occupy the ``F'' region \citep{Boyer+08,Boyer+09a,McDonald+2012}. The
D15 models support our assertion that the a-AGB are low mass stars
with insufficient dredge-up: those with the lowest masses will remain
O-rich, while those with higher masses will produce significant
quantities of O-rich dust prior to becoming C stars.

At lower metallicities, the lower-mass limit for the O- to C-rich
transition decreases, and the D15 models predict that the F sequence
should therefore be more sparsely populated in metal-poor
environments. Indeed, the ``F'' finger is not pronounced in the more
metal-poor SMC (Fig.~\ref{fig:dell1}). This is partially caused by
limited sensitivity at 24~\micron, which is 0.4~mag worse in the more
distant SMC.  However, since the a-AGB stars increase in brightness
at 24~\micron\ as color increases, any in the ``F'' region of the CMD
in the SMC are detectable. We thus conclude that metallicity is the
dominant cause of the empty ``F'' region in the SMC. This effect may
be compounded by less-efficient dust production in O-rich stars at low
metallicity due to a lack of condensation seeds, which will result in
fewer O-rich stars with 24-\micron\ excess. Based on the $Z=0.004$ D15
models, any low-mass stars in the SMC that will remain O-rich must
have initial masses $<1.25~M_\odot$, in good agreement with the masses
inferred from the star-formation histories.

The Padova TP-AGB
models\footnotemark[1]\footnotetext[1]{http://stev.oapd.inaf.it/cgi-bin/cmd}
\citep{Marigo+08} generally follow the D15 models, with slightly
higher mass transitions.  Here, we use the option with 60\% silicate
$+$ 40\% AlO$_{\rm x}$ for O-rich stars and 85\% amorphous carbon $+$
15\% SiC for C-rich stars \citep{Groenewegen06}.  These models show
that, in the LMC, all stars with initial mass $<$1.5~$M_\odot$ remain
on the ``F'' branch marked in Figure~\ref{fig:dell1}; transitions over
to the C-star side (left side) of the diagram occur only for stars
with $>$1.6$M_\odot$.  The Padova models indicate that this transition
happens at $M_{\rm i} \approx 1.3~M_\odot$ in the SMC.

\subsection{Probing the lower mass bound for C stars}

The a-AGB photometric classification appears to include initial
stellar masses that straddle the lower-mass limit for the O-to-C
transition caused by the dredge-up of carbon. They therefore include
both stars that will undergo this transition and stars that will
remain O-rich.  We can probe this transition several ways using the
a-AGB stars.

First, this boundary is reflected in the pulsation masses estimated in
Section~\ref{sec:param} for stars pulsating in the fundamental mode
(Sequence~1). The median pulsation mass for a-AGB stars is
1.14~$M_\odot$ in the LMC and 0.94~$M_\odot$ in the SMC, with
dispersions of 0.21 and 0.18~$M_\odot$, respectively
(Section~\ref{sec:param}).

Second, the luminosities of S-type stars can probe the O-to-C mass
transition. In Figure~\ref{fig:ao_pl}, we show the spectral
classifications of a-AGB stars on the 3.6-\micron\ $P$-$L$ diagram. In
the SMC, we see a curious pile-up of S-type a-AGB stars on Sequence 2
near $[3.6]=11.3$--$11.8$~mag, corresponding to $\log
(L/L_\odot)=3.6$--$3.7$ and $-7.6 < M_{3.6} < -7.1$~mag. Assuming
these are all intrinsic S-type stars (Section~\ref{sec:pulse}), it
suggests that stars fainter than this cutoff on Sequence~2 will likely
remain O-rich as they evolve to Sequence~1. Indeed, this is supported
by the relative lack of faint C-rich a-AGB stars on Sequence~1 in both
galaxies. This scenario would be confirmed if LMC S-type stars showed
a similar S-type star limit, but at a brighter luminosity.
Unfortunately, the only $s$-process molecular feature covered by the
LMC spectra is LaO, which is too weak for reliable S-type
classification.

\begin{figure}
  \includegraphics[width=\columnwidth]{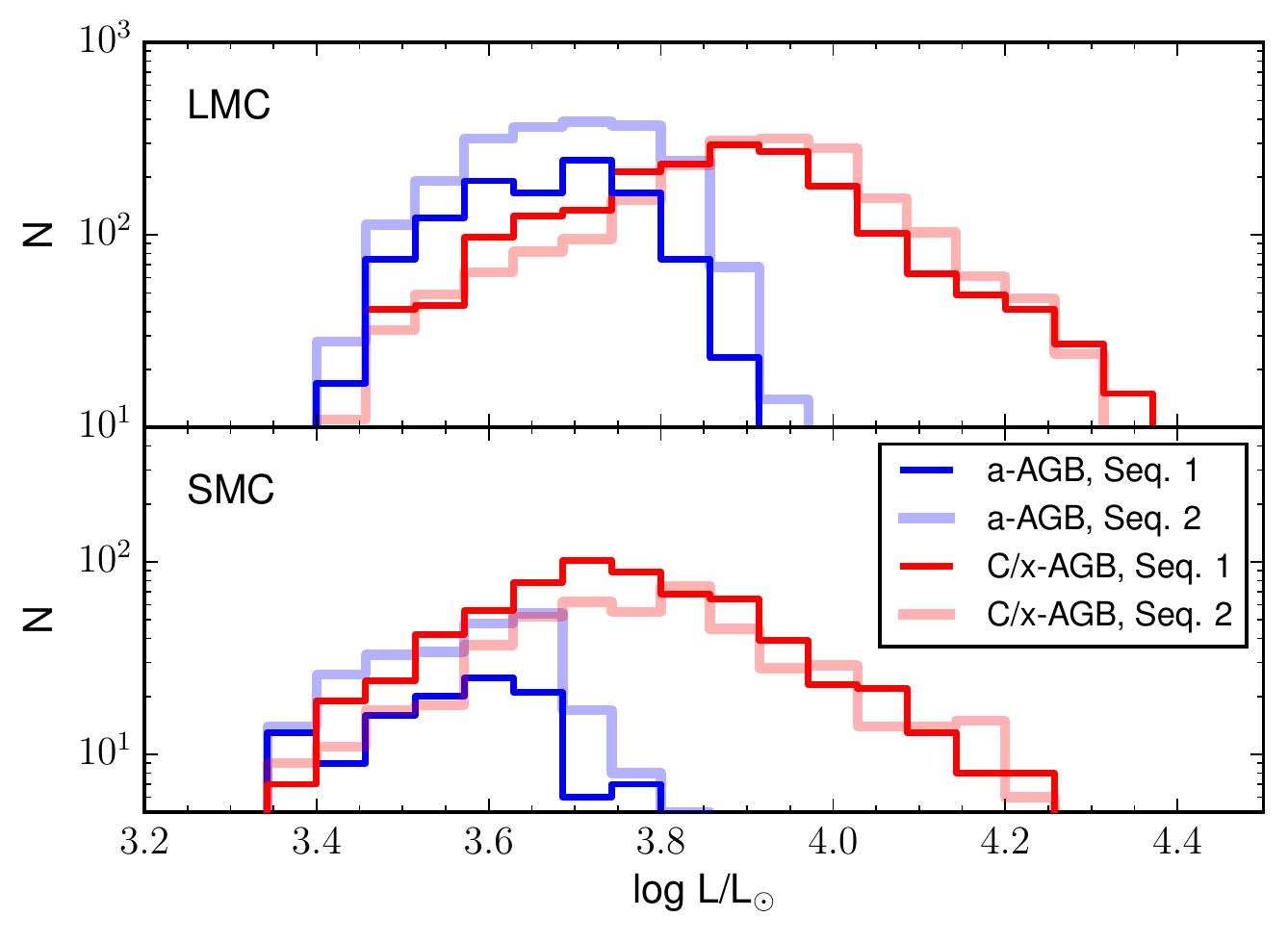}
  \figcaption{Luminosity functions for a-AGB stars and
    photometrically-classified C-rich stars (including C-AGB and x-AGB
    stars). We used the spectral classifications to minimize
    contamination from O-rich sources in the C-rich luminosity
    function. The increase in C-rich stars and decrease in a-AGB
    stars from Sequence~2 to 1 over the same luminosity range suggests
    that some Sequence~1 C-AGB stars originate from the Sequence~2
    a-AGB sample. \label{fig:lfunc}}
\end{figure}

Third, a comparison of the luminosity functions of a-AGB stars and
C-rich AGB stars on different pulsation sequences could be used to
calibrate the O-to-C transition in stellar evolution models. We show
these functions in Figure~\ref{fig:lfunc}, combining the C-AGB and
x-AGB photometric classes into the C-rich category. In both galaxies,
the number of a-AGB stars decreases from Sequence~2 to 1, and the
number of C-rich stars increases in the same luminosity range, hinting
that these new C-AGB stars originated from the Sequence~2 a-AGB
sample. Moreover, the change in the ratio of a-AGB stars to C-rich
stars from Sequences~2 to 1 is larger in the SMC by a factor of 1.4,
indicating a higher transition efficiency. This is consistent with
stellar evolution models, which predict a higher dredge-up efficiency
at low metallicities due both to less available free oxygen and
because the depth of the dredge-up events increases at low
metallicities.

Stellar evolution models
\citep[e.g.,][]{Karakas+02,Marigo+08,Marigo+13} also predict that the
dredge-up efficiency rapidly decreases at lower masses. The late
transition from O- to C-rich among the a-AGB stars supports this
prediction, suggesting that they require more dredge-up events to
reach ${\rm C/O}>1$ compared to their more massive C-AGB
counterparts. \citet{Karakas2014} compute that C/O in an LMC-like
1.5~$M_\odot$ star ($Z=0.007$) will increase by $\approx$0.2--0.25
after each thermal pulse. With an initial C/O near 0.4, LMC a-AGB
stars thus require a few thermal pulses to become C stars. This is
reflected in the low fraction (23\%) of C-rich a-AGB stars in the
LMC.

According to the same \citet{Karakas2014} models, an SMC-like star
($Z=0.004$, $M=1.5~M_\odot$), becomes C-rich after a single pulse,
with an increase in C/O of 1.5.  We would therefore expect about half
of the a-AGB stars to be C-rich if they are all at or above the
O-to-C transition mass, and indeed this is exactly what we find.  This
indicates that there are few stars in the SMC with masses below the
O-to-C transition that have reached the final stage of the AGB phase.

\subsection{Long Secondary Periods}
\label{sec:seqD}

Stars with long secondary periods (LSPs) lie on pulsation Sequence~D,
the cause of which is unknown.  With the collection of data we have
here, we have found some clues to the origin of the LSP: {\it (1)}
Sequence~D is dominated by O- and a-AGB stars (80\%; Table~\ref{tab:PL});
{\it (2)} DPRs on Sequence~D are similar to those on Sequences 1 and 2;
{\it (3)} unlike on other Sequences, the ratio of C/M stars on Sequence~D
is almost the same in both galaxies (0.39 in the LMC; 0.46 in the
SMC), despite the difference in metallicity; and {\it (4)} there are
almost no x-AGB stars on Sequence D.

The lack of x-AGB stars on Sequence~D may be partially caused by
circumstellar extinction; a significant fraction are not detected as
variable in OGLE, but are recovered as variable in the infrared
\citep{Riebel+2015}. If this is the case, then any x-AGB stars with
LSPs are dustier than those found on Sequence~1.  Of the a-AGB stars
that show significant dust production (based on whether they are
detected at 24~\micron), 19\% (LMC) and 32\% (SMC) are found on
Sequence~D. This is consistent with \citet{Wood+09}, who find that the
LSP is associated with dusty mass ejection.

\citet{Wood2015} show that the primary periods (i.e., the
second-longest period) of most LSP stars lies near Sequences 2 and 3,
as expected if most of the LSP stars are O- or a-AGB.  The preference
for stars with low initial masses and O-rich chemistries on Sequence~D
is a compelling hint for the nature of the LSP and may point to a
connection between the LSP and the initial onset of dust production. A
higher fraction of the a-AGB stars reside on Sequence~D in the SMC,
suggesting that metallicity may also play a role in the origin of the
LSP.

\subsection{Implications for Photometric Classification}
\label{sec:aoclass}
\begin{figure}
\vbox{
  \includegraphics[width=\columnwidth]{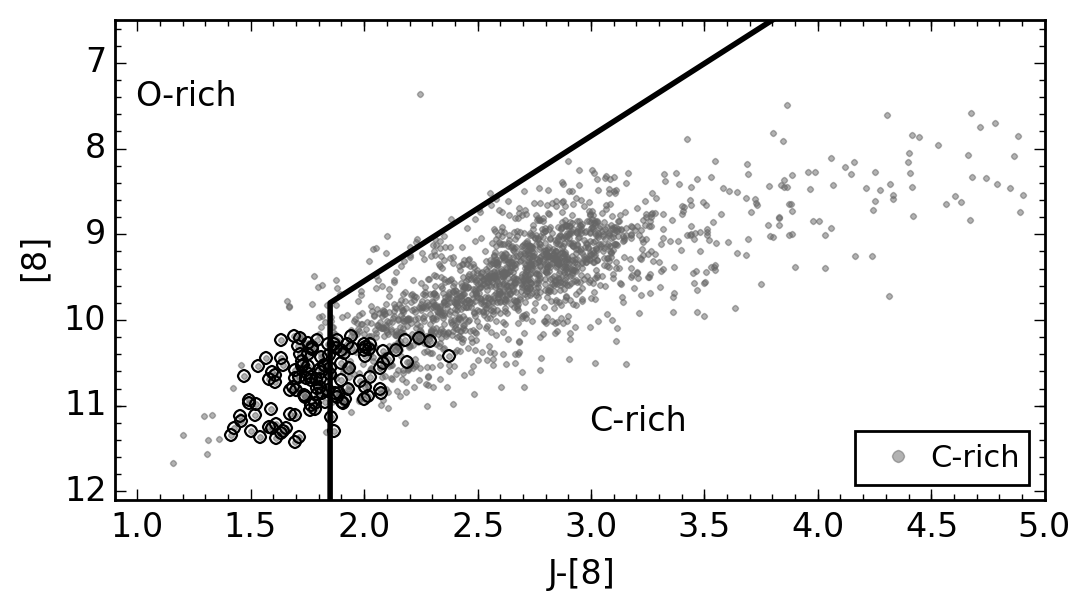}
  \includegraphics[width=\columnwidth]{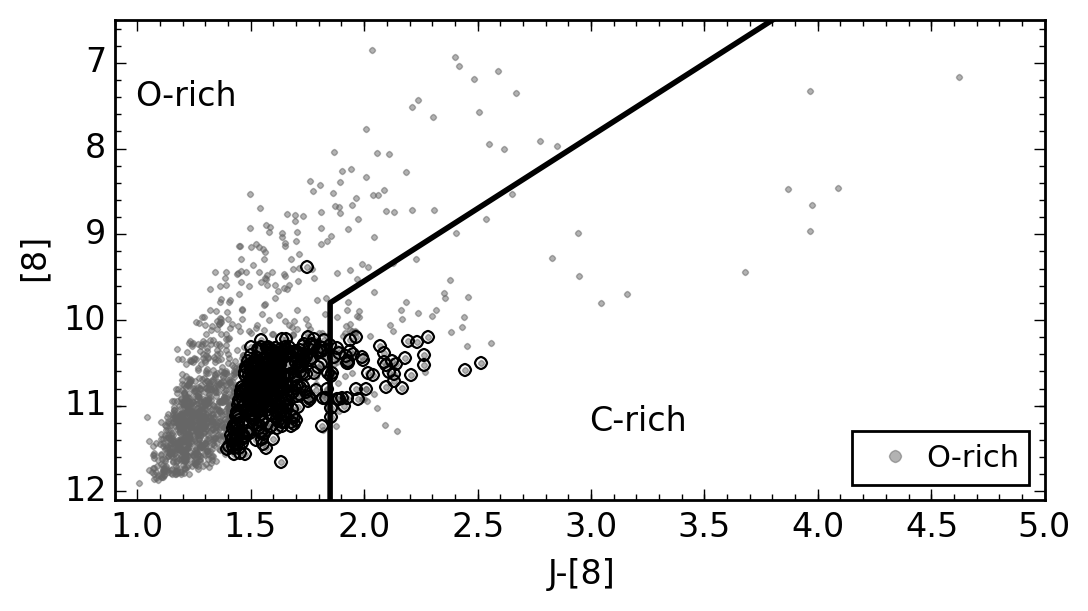}
} \figcaption{LMC Spectral classifications compared to the $J-[8.0]$
  color. The upper and lower panels show the spectrally-classified
  C-rich and O-rich stars, respectively. Points outlined in black were
  photometrically classified as a-AGB stars. The solid line marks the
  best photometric division between C-rich and O-rich stars.
\label{fig:jkclass}}
\end{figure}

\begin{figure*}
  \includegraphics[width=\textwidth]{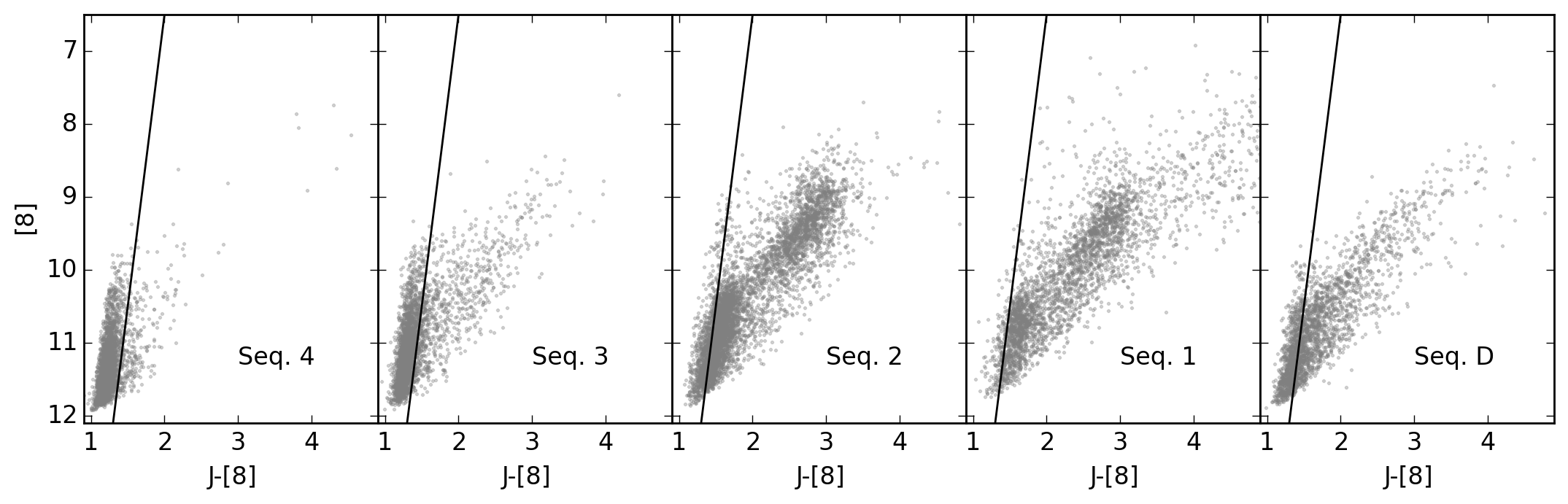}
  \figcaption{Distribution in $J-[8]$ color for stars on each
    pulsation sequence. The most evolved stars, particularly the
    low-mass (a-AGB) stars, can be identified as those redder than
    the solid line that intersects the upper and lower axes at
    $J-[8]=1.3$~mag and $J-[8]=3$~mag. \vspace{1em}
\label{fig:j8pulse}}
\end{figure*}

Even though the a-AGB stars have distinct $J-[8]$ colors from the
O-AGB and C-AGB populations, the \citet{Cioni+06b} $J-K_{\rm S}$
classification scheme works relatively well in the MCs from a
statistical perspective (Section~\ref{sec:specclass}).  However, this
scheme fails when the goal is to classify the most evolved,
dust-producing sources, including both the a-AGB stars and the more
massive O-AGB stars with high DPRs (Fig.~\ref{fig:jkclass}). The
results for M31 described in \citet{Boyer+2013} suggest that the
fraction of misclassified O-rich stars will also increase with
metallicity due to strong water absorption in late-type O-rich giants.

The black line in the lower panel of Figure~\ref{fig:jkclass} shows
the best $J-[8]$ separation in the LMC, resulting in the correct
classification for 93\% of the C stars and 92\% of the O-rich stars.
The misclassifications may be partially explained by non-simultaneous
observations between 2MASS and {\it Spitzer}, which can cause
artificially blue or red colors owing to variability. The {\it James
  Webb Space Telescope} (JWST) will include filters covering
0.7--25.5~\micron, allowing for near-simultaneous imaging over a wide
wavelength range. JWST will also have a filter centered on the
10-\micron\ silicate feature present in most dusty O-rich stars, which
may be a better choice than the IRAC 8.0-\micron\ filter for
separating dusty M and C stars.

Even though the statistical separation of C and M stars is only
marginally improved by including the 8-\micron\ photometry, we
recommend using the $J-[8]$ color to distinguish {\it evolved} TP-AGB
stars (especially the a-AGB stars) from their less evolved
counterparts when pulsation information is not available. This
separation is impossible in $J-K_{\rm S}$ and in various combinations
with the other three IRAC filters. We show the $J-[8]$ color distribution
for stars on each pulsation sequence in Figure~\ref{fig:j8pulse};
there is overlap depending on the exact definition of each pulsation
sequence, but it is clear that stars redder than the solid line are
dominated by evolved AGB stars on sequences 1 and 2 (a-AGB, C-AGB,
x-AGB, and massive O-AGB stars). Without this or a similar color cut,
it is difficult to identify the highly evolved, low-mass stars (the
a-AGB stars) without either the pulsation information or complete
spectral energy distribution fits for estimating the DPR.

The $[8]-[24]$ color (Fig.~\ref{fig:dell1}) is also a reliable way to
separate the evolved a-AGB stars from less-evolved O-AGB
stars. However, it results in an incomplete sample owing to the
limited sensitivity at long wavelengths. JWST will be approximately
3--4 times more sensitive than {\it Spitzer} at 24~\micron.

\subsection{The Cause of the Red $J-[8]$ colors}
\label{sec:cause}

\begin{figure}
  \includegraphics[width=\columnwidth]{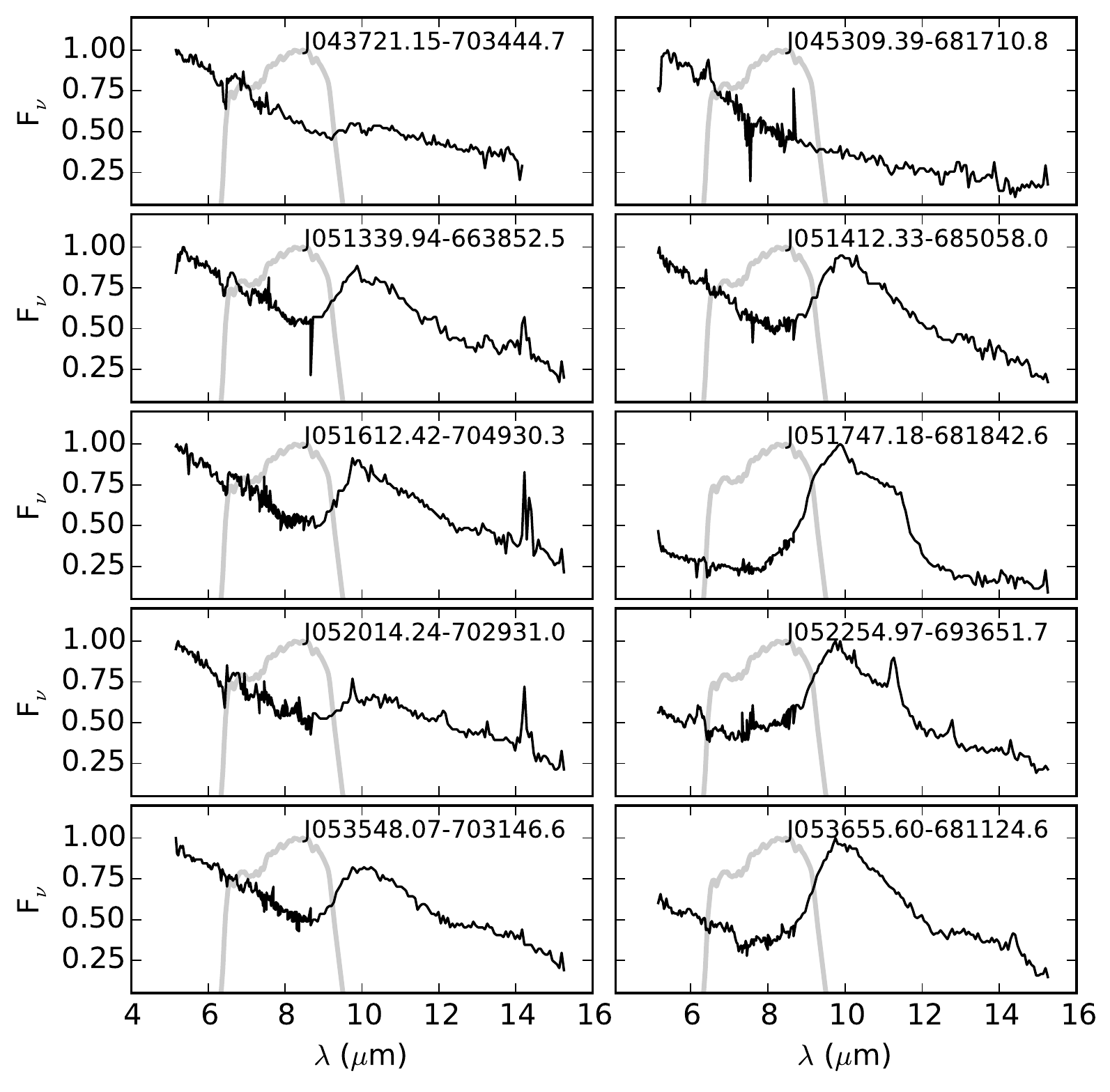} \figcaption{IRS
    spectra of a-AGB stars in the LMC \citep{Kemper+10}, in normalized
    flux density.  Nine show silicate emission at 10~\micron. The
    8-\micron\ IRAC filter (grey line) clips the wing of the silicate
    feature, causing red $J-[8]$ colors. The stars with the strongest
    silicate features (in the lower three panels on the right) show
    $J-[8]>2$~mag.\label{fig:irs}}
\end{figure}

B11 first noticed the a-AGB stars by their red $J-[8]$ colors. Their
colors must be caused by the flux in the 8-\micron\ filter since the
a-AGB stars do not stand out in other color combinations involving
$J$.  There are two possible spectral features that might affect this
wavelength range: SiO absorption at 8~\micron\ and silicate dust
emission at 10~\micron.  The a-AGB stars have modestly high DPRs,
confirming the presence of dust. Dust can cause both strong silicate
emission and/or veiling of the SiO molecular band, and both scenarios
would increase the $J-[8]$ color.  Ten of the LMC a-AGB stars have
been observed with the IRS and their AGB nature confirmed. Of these,
nine show strong silicate emission (Fig.~\ref{fig:irs}).  

In dustless
stars, \citet{Sloan+08} showed that the strength of SiO absorption
decreases with metallicity. If SiO is the dominant factor causing the
red $J-[8]$ color, we would therefore expect redder $J-[8]$ colors for
the SMC a-AGB stars.  Figure~\ref{fig:j8hist} shows the distribution
in a-AGB colors, and the SMC colors are, in fact, slightly {\it
  bluer}, on average.  We therefore conclude that the silicate feature
is the dominant cause of the red $J-[8]$ color for O-rich a-AGB
stars.  The $J-[8]$ colors of the massive dusty O-AGB stars are
similar to the a-AGB colors (see, for example,
Fig.~\ref{fig:dprsum}).  It is therefore reasonable that the red
colors are caused by the same mechanism, i.e., the silicate feature. 

B11 speculated that the red $J-[8]$ colors of a-AGB stars might be
caused by an unusually shaped silicate feature. The silicate features
of these nine examples are not remarkable compared to other O-rich AGB
stars in the SAGE-Spec program \citep{Woods+2011,Ruffle+2015}, so we
rule out an unusual feature shape as a cause of the red colors of
a-AGB stars. Instead, the red colors are more likely caused by the
overlap between the wing of the silicate feature and the
8-\micron\ filter transmission (Fig.~\ref{fig:irs}).

Dusty C-rich a-AGB stars would also be red in $J-[8]$ because of the
contribution of the dust continuum of amorphous carbon dust at
8~\micron; dustier stars would exhibit redder colors. Compared to
other C-AGB stars, the C-rich a-AGB stars have low DPRs, so their
colors at the blue end of the C-AGB branch in Fig.~\ref{fig:photclass}
is to be expected. 

C-rich and O-rich a-AGB stars are similarly indistinguishable in
color combinations that include the AKARI 11-\micron\ flux since both
C-rich and O-rich dusty stars have strong excess at 11~\micron.

\section{Conclusions}

Using {\it Spitzer} SAGE data, \citet{Boyer+11} noted a group of
thermally-pulsing AGB stars in the Magellanic Clouds which classify as
O-rich based on their $J-K_{\rm S}$ colors, but that show redder
$J-[8]$ colors than the bulk of their parent population. That work
dubbed these the anomalous O-rich (aO-)AGB stars.  We have gathered
evidence from multiple perspectives to ascertain the nature of these
stars and conclude that they are low-mass AGB stars that are
straddling the mass limit where dredge up is sufficient to transform
them into carbon stars. Since a high fraction of the aO-AGB stars
turned out to be C-rich, we now refer to the B11 aO-stars as simply
a-AGB stars, and find that they are evolved, low-mass dusty AGB
stars. We reach this conclusion by interpreting the following
observations:

\begin{itemize}

\item Their optical spectra show a high fraction of both C- and O-rich
  chemistries despite having $J-K_{\rm S}$ colors that are typically
  assigned to O-rich stars. There is a higher fraction of C-rich
  a-AGB stars in the more metal-poor SMC (50\%, compared to 23\% in
  the LMC).

\item Light curves indicate they are pulsating primarily in the
  fundamental mode and first overtone, suggesting they are at the very
  end of their evolution.  Among those pulsating in the fundamental
  mode, the a-AGB median pulsation masses are lower than those of
  the O-AGB and C-AGB populations.

\item We identified (likely intrinsic) S-type stars in the SMC, and
  find that they comprise 9\% of the a-AGB population.  These pulse
  primarily in the first overtone and cluster near
  $[3.6]=11.3$--$11.8$~mag, suggesting that stars with similar
  properties will transition to C-rich as they move to the fundamental
  mode.

\item Their dust-production rates are consistent with low-mass O-AGB
  stars.

\item Their colors match stellar evolution tracks for low-mass
  AGB stars.

\item Their metallicities are low compared to the O-AGB stars,
  implying that they originated from an older, less
  chemically-evolved, stellar population.

\end{itemize}

The results here suggest that a-AGB stars can be easily selected
photometrically using the $J-[8]$ color, which is more reliable than
$J-K_{\rm S}$ for the most evolved stars, and particularly for the
dust-enshrouded O-rich stars. In addition, surveys searching for
S-type stars may be wise to search among the brightest a-AGB stars. 

With the compiled rich data set for the Magellanic Clouds, we have
noted some general properties of the a-AGB stars in addition to the
evidence listed above. First, their dust-production rates are not
strongly dependent on their periods, at least compared to other O- and
C-AGB stars pulsating in the fundamental mode (Mira
variables). Second, they are over represented among stars with a long
secondary period. Third, their effective temperatures are similar to
high-luminosity O-rich evolved stars. Finally, even those that will
eventually transition into C-AGB stars will produce significant
quantities of O-rich dust prior to this transition. More massive stars
will become C-rich before the onset of strong dust production (except
the most massive hot-bottom-burning AGB stars, which will remain
O-rich).

In the LMC, models suggest the a-AGB stars have initial masses up to
$1.5~M_\odot$, and up to $1.25~M_\odot$ in the SMC.  The transition
mass would therefore be a bit below this limit. In both of the
Magellanic Clouds, there is a peak in the star-formation history that
corresponds to initial stellar masses of 1--1.3~$M_\odot$
\citep{Harris+04,HarrisZaritsky2009}. The evidence presented here
suggests that this star formation event is the origin of the a-AGB
stars, and may have been caused by an encounter between the LMC and
SMC $\approx$2.5~Gyr ago.

\acknowledgements

We thank Paolo Ventura \& Flavia Dell'Agli for providing their stellar
evolution tracks, and Amanda Karakas for providing dredge-up
models. We also thank Patricia Whitelock for helpful discussions on
stellar variability.  This research made use of NASA's Astrophysics
Data System; the NASA/IPAC Infrared Science Archive, which is operated
by JPL/California Institute of Technology, under contract with the
NASA; the SIMBAD database, operated at CDS, Strasbourg, France;
Astropy, a community-developed core Python package for Astronomy
\citep{astropy}. This publication also makes use of data products from
the Two Micron All Sky Survey, which is a joint project of the
University of Massachusetts and IPAC/California Institute of
Technology, funded by NASA and the NSF; and the Wide-field Infrared
Survey Explorer, which is a joint project of the University of
California, and the JPL/California Institute of Technology, funded by
NASA. MLB is supported by the NASA Postdoctoral Program at the Goddard
Space Flight Center, administered by ORAU through a contract with
NASA.

\bibliographystyle{astroads}
\bibliography{myrefs}

\end{document}

%% file: abstract_v2.tex
We have identified a new class of Asymptotic Giant Branch (AGB) stars
in the Small and Large Magellanic Clouds (SMC/LMC) using optical to
infrared photometry, lightcurves, and optical spectroscopy. The strong
dust production and long-period pulsations of these stars indicate
that they are at the very end of their AGB
evolution. Period-mass-radius relations for the fundamental-mode
pulsators give median current stellar masses of $1.14~M_\odot$ in the
LMC and $0.94~M_\odot$ in the SMC (with dispersions of 0.21 and
0.18~$M_\odot$, respectively), and models suggest initial masses of
$<$1.5 $M_\odot$ and $<$1.25 $M_\odot$, respectively. This new class
of stars includes both O-rich and C-rich chemistries, placing the
limit where dredge-up allows carbon star production below these
masses. A high fraction of the brightest among them should show S star
characteristics indicative of atmospheric C/O $\approx$ 1, and many
will form O-rich dust prior to their C-rich phase. These stars can be
separated from their less-evolved counterparts by their
characteristically red $J-[8]$ colors.